\documentclass[twocolumn,showkeys,nofootinbib,preprintnumbers,amsmath,amssymb,prd,floatfix]{revtex4}

\usepackage{graphicx}
\usepackage{dcolumn}
\usepackage{bm}
\usepackage{natbib}
\usepackage{verbatim}
\usepackage{latexsym}

\begin{document}
\title{RADIATION FROM BODIES WITH EXTREME ACCELERATION II:
KINEMATICS\footnote{With hearty felicitations to Jacob Bekenstein, the
father of black hole entropy.}\footnote{Published in ``Thirty Years of
Black Hole Entropy'', a special issue, \emph{Foundations of Physics}
\textbf{33}, pp 179-221 (2003); \url{http://www.math.ohio-state.edu/~gerlach} } }


\author{ULRICH H. GERLACH}
\email{gerlach@math.ohio-state.edu}
\affiliation{%
Department of Mathematics, Ohio State University, Columbus, OH 43210, USA
}%

\date{\today}

\begin{abstract}
When applied to a dipole source subjected to acceleration which is
violent and long lasting (``extreme acceleration''), Maxwell's
equations predict radiative power which augments Larmor's classical
radiation formula by a nontrivial amount. The physical assumptions
behind this result are made possible by the kinematics of a system of
geometrical clocks whose tickings are controlled by cavities which are
expanding inertially. For the purpose of measuring the radiation from
such a source we take advantage of the physical validity of a
spacetime coordinate framework (``inertially expanding frame'') based
on such clocks. They are compatible and commensurable with the
accelerated clocks of the accelerated source.  By contrast, a common
Lorentz frame with its mutually static clocks won't do: it lacks that
commensurability. Inertially expanding clocks give a physicist a
window into the frame of a source with extreme acceleration. He thus
can locate that source and measure radiation from it without being
subjected to such acceleration himself. The conclusion is that
inertially expanding reference frames reveal qualitatively distinct
aspects of nature which would not be accessible if static inertial
frames were the only admissible frames.
\end{abstract}
\maketitle
\tableofcontents
\section{INTRODUCTION}
Minkowski spacetime with its Lorentz geometry is the geometrical
framework for most physical measurements, in particular those
involving radiation and scattering processes. Indeed, the asymptotic
``in'' and ``out'' regions of the scattering matrix, as well as the
asymptotic ``far-field'' regions of a radiator reflect this fact. 

If
the scatterer or radiation source is accelerated linearly and uniformly, then
the standard approach is to characterize its coaccelerating coordinate
frame in terms of a one-parameter family of instantaneous Lorentz
frames, any one of which provides the necessary ``in'' and ``out'' or
``far-field'' regions for the measurements of the scattered and
emitted radiation.

However, suppose the acceleration is \emph{extreme}, i.e. 
\begin{itemize}
\item
the acceleration, say $g$, lasts long enough for the
scatterering/radiation source to acquire relativistic velocities and
\item
is large enough to do this within one cycle of its characteristic
frequency $\omega$ so that
\[
\frac{g}{\omega} \sim \textrm{(speed of light)}~.
\]
\end{itemize}
Under such a circumstance no physicist who insists on using a given
asymptotic Lorentz frame as his observation platform can escape from a
number of difficulties trying to execute his measurements.

First of all, there is the distortion problem. Relative to any Lorentz
frame a signal emitted by, say, an accelerated dipole would be
subjected to a time-dependent Doppler shift (``Doppler chirp''). The
received signal starts out with an extreme blue shift and finishes
with an extreme red shift. Such a distortion prevails not only in the
time domain, but also in the spatial domain of that Lorentz
frame. Once this distorted signal has been acquired by the observer in
his Lorentz frame, he is confronted with the task of applying a time
and/or space transformation to remove this distortion. He must
reconstruct the signal in order to recover with 100\% fidelity the
original signal emitted by the source. Such a task is tantamount to
changing from his familiar set of clocks and meter rods, which make up
his Lorentz frame, to a new set of clocks and units of length relative
to which the signal presents itself in undistorted form with 100\%
fidelity.

Second, there is the problem of the trajectory of the accelerated
source. In order to have a ``far field'' region, the source must be
much smaller than one wavelength. If the acceleration lasts long
enough, the source will reach within one oscillation the
asymptotically distant observer where the measuring equipment is
located and thus vitiate its status as being located in the ``far
field'' region: there no longer is large sphere that surrounds the
source\footnote{This difficulty might not be of much bother to the
physicist who can find sources which are subject to extreme
acceleration but which cease to exist well before they reach him.}.

Finally, during such an acceleration process the source would be
emitting plenty of information about itself (in the form of spectral
power, angular distribution, etc.). However, to acquire this
information the Maxwell field must be measured in the radiation zone.
It lies outside a sphere centered around the source with radius one
wave length. (Inside this sphere the radiation field is inextricably
mixed up with the ``induction'' field.) Measuring the Maxwell field
consists of relating its measured amplitude to the synchronized clocks
and measuring rods. But this is precisely what cannot be done if the
wavelength is larger than (acceleration)$^{-1}$ of the accelerated
source. In that case the far field falls outside the semi-infinite
domain \cite{MTW:6.3} where the events are characterized uniquely by
the clocks that are synchronized with the accelerated clock of the
source. Put differently, the semi-finite size of the ``local
coordinate system of the accelerated source'' \cite{MTW:6.6} does not
allow an observer to distance himself far enough from the source to
identify the radiative field in the far zone.

Aside from removing the above ambiguities, the purpose of this note is
to identify the spacetime framework which accommodates Maxwell's field
equations applied to a uniformly and linearly accelerated radiation
source.  One such application is the radiation observed in response to
a dipole source.  The observed radiation rate is given by the familiar
Larmor formula but augmented due to the unique source-induced
spacetime framework.  This enhanced Larmor radiation formula is the
result of a straightforward calculation based on this
framework. There are no arbitrary hypotheses. The formula is given by
\cite{Gerlach:2001}:
\begin{equation}
\left(
\begin{array}{c}
   \textrm{flow of radiant}\\
   \textrm{energy into the}\\
   \textrm{direction of acc'n}
\end{array}
\right)
=\frac{2}{3}\left[
\left(
\frac{d^2 D}{d\tau^2} \right)^2
+
\left(
\frac{d D}{d\tau} \right)^2
            \right]
\label{eq:geometrical Larmor formula}
\end{equation}
Here 
\[
D=
\left(
  \begin{array}{c}
   \textrm{proper}\\
   \textrm{dipole}\\
   \textrm{moment}
  \end{array}
  \right) \times
\left(
  \begin{array}{c}
   \textrm{proper acc'n}\\
   \textrm{of the}\\
   \textrm{dipole}
  \end{array}
  \right)
\]
is the \emph{geometrical dipole moment}. It
is the time dependent magnitude of a dipole source pointing along the
direction of acceleration, which is linear and uniform. Furthermore,
\[
\tau= \left( \textrm{geometrical (dimensionless, ``Rindler'') time} \right) ~,
\]
and the quantity
\[
\left(
\begin{array}{c}
   \textrm{flow of radiant}\\
   \textrm{energy into the}\\
   \textrm{direction of acc'n}
\end{array}
\right)
=
\left(
\begin{array}{c}
  \textrm{linear density of}\\
  \textrm{longitudinal component}\\
  \textrm{of radiated momentum}
\end{array}
\right)~.
\]
is defined by the conserved $\tau$-momentum in boost-invariant sector $F$,
\begin{eqnarray*}
&\displaystyle\int_{-\infty}^\infty d\tau& 
\overbrace{\int_{0}^\infty \int_0^{2\pi}
T_{~\tau}^\xi \xi\,rdr d\theta}^{\textrm{evaluated at }\xi=const. \textrm{ in }F}=\\
&\displaystyle\int_{-\infty}^\infty d\tau& 
\left(
\begin{array}{c}
  \textrm{linear density of}\\
  \textrm{longitudinal component}\\
  \textrm{of radiated momentum}
\end{array}
\right).
\end{eqnarray*}
Formula (\ref{eq:geometrical Larmor formula}) expresses a causal link
between what happens at the accelerated source and the radiant energy
observed on the other side of event horizon. The $\tau$-coordinate is
the key.  It is a symmetry trajectory on both sides of this horizon.
This enables it to serve as the same standard for reckoning changes in
the source in $I$ as for reckoning changes in the location in $F$.

What is the spacetime framework, i.e. the nature of the arrangement of
measuring rods and clocks which makes this formula possible?
Even if the spacetime framework for the accelerated dipole source is
clear, comprehending Eq.(\ref{eq:geometrical Larmor formula}) entails
asking: (i) What is the spacetime framework for the observer who measures
the radiation? (ii) What is the relationship between his framework and
that of the source?  

\begin{figure}
\includegraphics[scale=.8,bb=100 440 600
720]{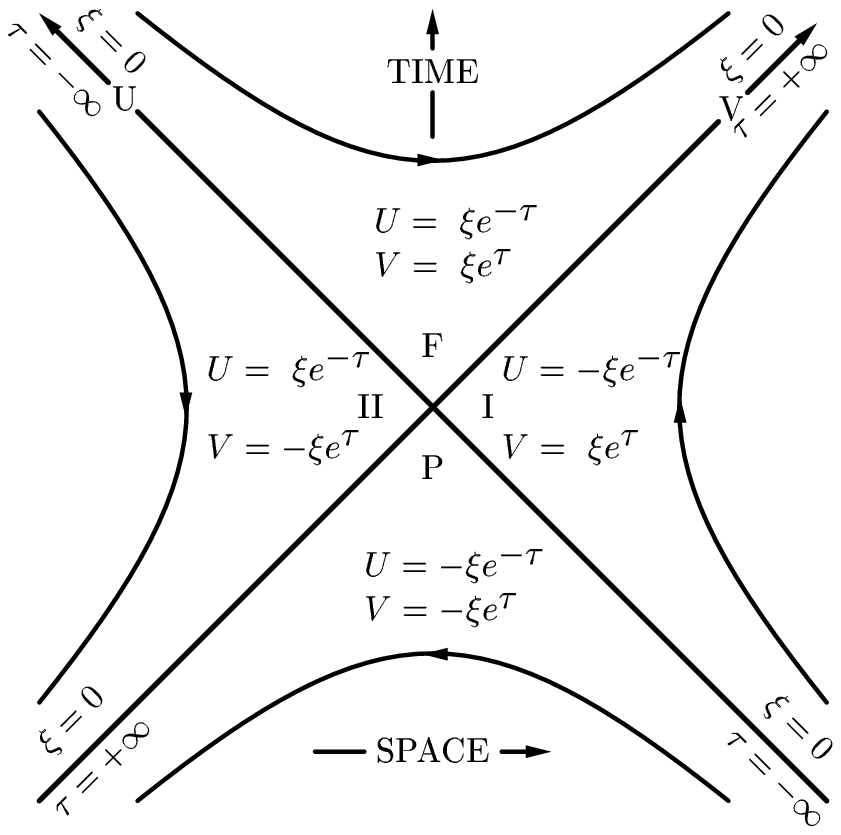}
\caption{\label{fig:Rindler spacetime} Acceleration-induced
partitioning of spacetime into four boost-invariant\protect\footnote{As far as can
be ascertained, the four acceleration-induced sectors $I,F,II,P$ and their
respective coordinates were first introduced by Bondi in \cite{Bondi}
and independently by Rindler in \cite{Rindler:1966}} 
sectors. They are
centered around the reference event $(t_0,z_0)$ so that
$U=(t-t_0)-(z-z_0)$ and $V=(t-t_0)+(z-z_0)$ are the retarded and
advanced time coordinates for this particular quartet of boost (a.k.a. ``Rindler'')
sectors.  The meaning of the boost coordinates $\xi$ and $\tau$ is
inferred from the expressions for the invariant interval $-dt^2+dz^2
=-\xi^2d\tau^2+d\xi^2$ in $I$ \& $II$ and
$-dt^2+dz^2=\xi^2d\tau^2-d\xi^2$ in $F$ \& $P$.  The emitted radiation
given by Eq.(\ref{eq:geometrical Larmor formula}) applies to a
point-like source whose world line traces out the hyperbola in sector
$I$.}
\end{figure}

As depicted in Figure \ref{fig:Rindler spacetime}, the source traces
out a world line which is hyperbolic relative to a globally free-float
observer, one with a system of inertial clocks in a state of relative
rest to one another. However, the observed energy given by
Eq.(\ref{eq:geometrical Larmor formula}) is to be measured by a
different observer, one whose clocks, even though also inertial,
have nonzero expansion relative to one another.

We shall find that the arrangement of clocks and rods of such an
observer is confined to future sector $F$ of Figure \ref{fig:Rindler
spacetime}. This sector is separated by the history of a one-way
membrane (``event horizon'') from the spacetime domain, sector $I$, of
the source. The purpose of this article is to establish a physical
bridge between the two, and bolt them together into a single arena
appropriate for the measurement of attributes of bodies subjected to
extreme acceleration, Eq.(\ref{eq:geometrical Larmor formula}) being
one of them.

The above questions do not deal with the inertia of bodies, nor with
the dynamics of material particles, nor with the dynamics of the
Maxwell field equations\footnote{Reference \cite{Gerlach:2001},
``Radiation from violently accelerated bodies'', dealt with the
dynamics of the Maxwell field equations, to which the present paper is
a sequel.}. Instead, they address kinematical aspects of the source
and the observer by introducing \emph{geometrical clocks} which are
\emph{commensurable}.

\section{GEOMETRICAL CLOCKS}
The measurement of radiation and other electrodynamical processes
depends on establishing a quantitative relationship ($x\rightarrow
\vec E(x),\vec B(x)$) between the electromagnetic field and a
coordinate reference system. Such a system consists of identically
constructed clocks, which, for a freely floating reference system
\cite{Taylor_and_Wheeler:1992}, are (i) synchronized and (ii)
separated by a standard unit (rigid meter rod) of length. However, 
measurements based on standard atomic clocks and on standard atomic
units of length have a number of disadvantages.

First of all, they do not take advantage of the fact that the speed of
light furnishes a unique and universal relation between the standard of
time and the standard of length. 

Second, for an accelerated reference system the (pseudo) gravitational
frequency shift frustrates the synchronization of clocks indicating
proper time.  

Third, the pseudo gravitational redshift changes the wavelength of
light and hence brings about discrepancies between the atomic standard
of length based on a fixed number of wavelengths of Krypton 86 (more
recently, of iodine stabilized He-Ne laser light) and the atomic
standard based on a fixed number of platinum atoms (rigid
platinum-iridium meter stick in Paris).

Last, but not least, there are regions which simply don't
lend themselves to being probed by rigid bodies, if for no other
reason than that the assumed rigidity of material meter rods loses its
meaning when the acceleration becomes high enough. 

The methods of Doppler radar and pulse radar do not suffer from any of
these disadvantages.  Moreover, it is possible to formulate the entire
kinematics of special relativity in terms of the methods of
radar\footnote{This has been done in a highly original way by Bondi in
\cite{Bondi}}.

\subsection{Radar Units}

Fundamentally, physics, including the kinematics of
moving bodies, is based on measurements. The class of measurements we
shall focus on are those made by identically constructed ``radar units'', 
each in its state of acceleration, which may be zero, in which case
it is in a state of ``free-float''. The meaning of ``radar units''
is that each one of them has
\begin{itemize}
\item
a Doppler radar which emits a standard frequency wave train,
controlled by an atomic clock of neglegible size (``atomic wrist
watch''),
\item
a frequency analyzer capable of measuring and recording the Fourier
spectrum of the reflected Doppler signal or of the pulse train
emitted by another radar unit,
\item
a reflecting surface so as to make the radar unit visible to the
radar of the other units, 
\item
a coincidence pulse radar which consists of a transponder
(receiver+transmitter), which upon receiving a pulse will emit without
any delay a replica of this pulse, and
\item
a recording device which counts and is capable of storing the intensity of the
received pulses.
\end{itemize}
The coincidence pulse capability implies that if there are two radar
units, then a pulse striking one of them initiates a process of a
pulse bouncing back and forth between the two radar units. This is
depicted in Figures~\ref{fig:inertial clock plus nullrays} and
\ref{fig:accelerated clock plus nullrays}, where the two radar units
are labelled A and B.

\subsection{Fourier Compatibility}

A fundamental property of every pair of radar units is their
compatibility with respect to frequency measurements. More precisesly,
one has the following definition:

Two radar units are said to be \emph{Fourier compatible} if
and only if a continuous wave train emitted by one radar unit produces
a return signal which has a sharp Fourier spectrum at the second radar
unit. If the return signal is not spectrally sharp (within
prespecified bounds), then the two units are said to be \emph{Fourier
incompatible}.

A pair of Fourier compatible radar units, say A and B, is
characterized by two frequency shift factors. The transmission process
A $\rightarrow$ B is characterized by
\begin{eqnarray*}
\lefteqn{ k_{AB} \equiv }\\ 
& &
\frac{\textrm{(frequency of atomic clock at B)}}{\textrm{(frequency of
atomic clock at A, but observed at B)}} ~,
\end{eqnarray*}
while the reverse transmission process B $\rightarrow$ A is
characterized by
\begin{eqnarray*}
\lefteqn{ k_{BA} \equiv }\\
& &
\frac{\textrm{(frequency of atomic clock at A)}}{\textrm{(frequency of
atomic clock at B, but observed at A)}}~.
\end{eqnarray*}

The numbers $k_{AB}$ and  $k_{BA}$ are, of course, the familiar
Doppler frequency shift factors if A and B are freely floating units,
and the pseudo-gravitational frequency shift factors if A and B
are uniformly and collinearly accelerated units. 
For the former one has
$k_{AB}=k_{BA}$, while for the latter one has $k_{AB}=1/k_{BA}$.

These frequency shifts (``Fourier compatibility factors'') are
strictly kinematical aspects of A and B.  They involve neither the
inertia nor the dynamics of material particles. Nevertheless, they do
distinguish between free-float and acceleration. Indeed this
distinction is encoded into the the relation between the frequency
shifts associated with the reflection process A $\rightarrow$ B
$\rightarrow$ A for monochromatic radiation.  There one has
\begin{eqnarray*}
k_{A\rightarrow B\rightarrow A}\equiv k_{A\rightarrow B}k_{B\rightarrow A}=
k_{AB}k_{BA}\\
~\\
=
\left\{
\begin{array}{cc}
  (k_{AB})^2 & 
           \begin{array}{c}
	     \textrm{whenever A $\&$ B}\\
	     \textrm{are floating freely}
	     \end{array}
	                           \\
~&~\\
  1          &
            \begin{array}{c}
	      \textrm{whenever A $\&$ B}\\
	      \textrm{are in states of}\\
	      \textrm{uniform collinear acc'n}
	    \end{array}  
\end{array}
\right.
\end{eqnarray*}


For example, it is clear that all freely floating (``inertial'')
units are Fourier compatible. So are the units which are linearly
and uniformly accelerated and have the same future and past event
horizons.  However, an accelerated unit and one in a state of free
float are not Fourier compatible. Neither are two units if one of
them undergoes non-uniform acceleration. Such units measure a Doppler
chirp instead of a constant Doppler shift when they receive the wave
train reflected by the other.

\subsection{Geometrical Clocks}

Time and space acquire their meaning from measurements,
i.e. identifications of a relationship by means of a unit that
serves as a standard of measurement. The measurement process we shall
focus on is based on the emission, reflection, and reception of radar
pulses generated by a standard \emph{geometrical clock}.

Such a clock consists of a pair of Fourier compatible radar units,
say, A and B. Their reflective surfaces form the two ends of a
one-dimensional cavity with its evenly spaced spectrum of allowed
standing wave modes in between. The operation of the geometrical clock
hinges on having an electromagnetic pulse travel back and forth
between the reflective ends of the cavity. The back and forth travel
rate is determined by the separation between the cavity ends. This
rate need not, of course, be constant in relation to the atomic clocks
carried at either end. A geometrical clock with mutually receding ends
would exemplify such a circumstance.

The definition of a geometrical clock is therefore this: it is a 
one-dimensional cavity 
\begin{itemize}
\item
whose bounding ends are Fourier compatible and
\item
which accommodates an electromagnetic pulse bouncing back and forth
between the left and right end of the cavity.
\end{itemize}
This bouncing action forms the tick-tock events of the clock. If the cavity
is expanding inertially, these events are located at
\begin{equation}
\begin{array}{clc}
   (t,z)=be^{n\Delta\tau}&(1,0)& \left( \begin{array}{c}
                                       n=\textrm{even}\\
				       \textrm{``tick''}
				  \end{array}\right) \\
   (t,z)=be^{(n+1)\Delta\tau}&(\cosh \Delta\tau,\sinh \Delta\tau)&  
                           \left( \begin{array}{c}
                                       n+1=\textrm{odd}\\
                                       \textrm{``tock''}
                                  \end{array} \right)
\end{array}
\label{eq:inertially expanding standard}
\end{equation}
along the two straight world lines of radar units A and B in spacetime 
sector $F$. Here 
\[
e^{\Delta\tau}=k_{AB}
\]
is the Doppler frequency shift factor and $\Delta\tau$ is the fixed
comoving separation between A and B. The constant $b$ is the Minkowski time
when the geometrical clock strikes zero.

For a clock with ends subjected to accelerations $1/b$ and
$1/be^{\Delta\tau}$, the ticking events are located at
\begin{widetext}
\begin{equation}
\begin{array}{clc}
   (t,z)=b&(\sinh n\Delta\tau, \cosh n\Delta\tau)& \left( \begin{array}{c}
                                       n=\textrm{even}\\
                                       \textrm{``tick''}
                                  \end{array}\right) \\
   (t,z)=be^{\Delta\tau}&(\sinh (n+1)\Delta\tau,\cosh (n+1)\Delta\tau)&
                           \left( \begin{array}{c}
                                       n+1=\textrm{odd}\\
                                       \textrm{``tock''}
                                  \end{array} \right)
\end{array}
\label{eq:accelerated standard}
\end{equation}
\end{widetext}
along the two hyperbolic world lines of A and B in boost-invariant sector $I$.
Here
\[
e^{\Delta\tau}=k_{AB}
\]
is the pseudo-gravitational frequency shift factor between them, and 
$\Delta\tau$ is the boost time between a ``tick'' and a ``tock''.

\begin{figure}
\includegraphics[scale=.5]{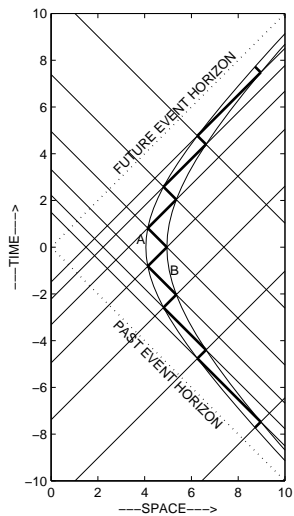}
\caption{\label{fig:accelerated clock plus nullrays} Spacetime history
of an accelerated clock and the null trajectories of trains of
emitted and received pulses (light lines) whose emission and reception is controlled by the
internal pulse (heavy line) bouncing back and forth between A and B.}
\end{figure}

The spacetime history of such clocks and their bouncing pulses are
exhibited in Figures \ref{fig:inertial clock plus
nullrays} and \ref{fig:accelerated clock plus nullrays}.

\begin{figure}
\includegraphics[scale=.5]{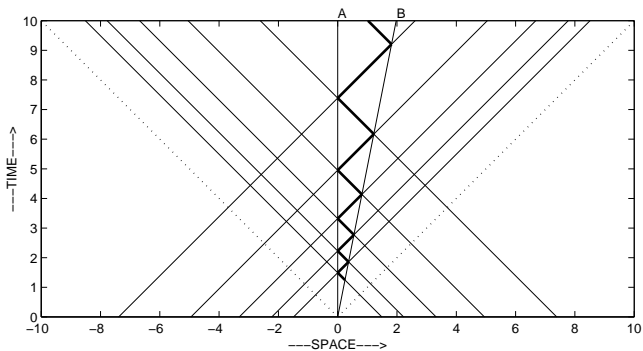}
\caption{\label{fig:inertial clock plus nullrays} Spacetime history of
an inertially expanding clock and the null trajectories of trains of
emitted and received pulses (light lines) whose emission and
reception is controlled by the internal pulse (heavy line) bouncing
back and forth between A and B.}
\end{figure}

To serve its purpose, a geometrical clock AB emits and receives pulses
of radiation. When the internal pulse strikes radar unit A its
transmitter and its receiver are turned on only for the duration of
the pulse. This causes A to emit a pulse and to register the reception
of a pulse from the outside, if there is one coming. When the internal
pulse has bounced back to B, an analogous emission and reception
process takes place at radar unit B. It follows that that the
tick-tock action of the internally bouncing clock pulse determines a
set of external pulses moving to the right and to the left. The
history of these pulses together with the clock that controls them is
depicted by Figures \ref{fig:inertial clock plus nullrays} and
\ref{fig:accelerated clock plus nullrays} for an inertially expanding and
accelerating clock respectively.

\section{PRINCIPLES OF MEASUREMENT}
\label{sec-Principles of Measurement}
Geometrical clocks play a fundamental role in the development of the
measurement of space and time. However, in order not to appear
arbitrary, following Rand\footnote{``Cognition and Measurement'',
Ch. 1 in \cite{Rand:1990}} and Peikoff\footnote{``Concept-Formation as
a Mathematical Process'', p.81 ff in \cite{Peikoff:1993}}, we shall
remind ourselves telegraphically of the nature of measurement from a
perspective which requires no specialized knowledge and no specialized
training.

A process of measurement involves two concretes: the thing being
measured and the thing that is the standard of measurement. The
relationship between the two is reciprocal: either one may serve as a
standard.  Measurements pertain to the attributes of these
concretes. The choice of one of them as a standard is based on having
its attribute serve as a unit of measurement. The process of
measurement consists of establishing a relationship to this unit which
serves as a standard of measurement.

Within certain limits the choice of a standard is optional.  However,
the primary standard must be in a form (e.g. platinum meter rod in
Paris, or Cesium clock at N.I.S.T. in Boulder, Colorado, etc.) easily
accessible to a physicist and it must represent the specific attribute
which serves as a unit of measurement (e.g. 1 meter of length, or 1
second = 9,192,631,770 Cesium cycles of time, etc.). Moreover,
\emph{once a standard has been chosen, it becomes immutable for all
subsequent measurements}.  Any chosen standard satisfies this
principle.  A standard gets copied in the form of secondary standards.
Their purpose is to establish -- usually by a process of counting -- a
quantitative relationship between the standard and any other instance
of the attribute of the thing to be measured.

Whenever certain concretes have attributes which can be related to the
same standard of measurement, one says that these concretes are
\emph{commensurable}. The importance of commensurability lies in the
fact that it is an equivalence relation: If concrete $\mathcal A$ is
commensurable with concrete $\mathcal B$, then $\mathcal B$ is commensurable
with $\mathcal A$; if $\mathcal A$ is commensurable with $\mathcal B$,
and $\mathcal B$ is commensurable with $\mathcal C$, then $\mathcal A$
is commensurable with $\mathcal C$. Using this fact, and omitting
explicit reference to the specific measurement of their attributes, but
retaining their existence, a physicist integrates these concretes into
an equivalence class. 

Thus, based on commensurability with a standard rod, one forms an
equivalence class, the concept \emph{length}. Or, based on
commensurability with an entity undergoing a periodic process, one
forms another equivalence class, the concept \emph{time}.

A century ago physicists thought that the concept of length and of
time required two independent standards, one for each. But in 1905 it
was realized that these two standards are not independent. In fact,
they are related by a universal conversion factor, the speed of light
in vacuum. Thus starting in 1983 both length and time have been
defined by referring to a single standard, a unit of time as
determined by the tickings of a Cesium atomic clock. 

\begin{figure}
\includegraphics[scale=.5]{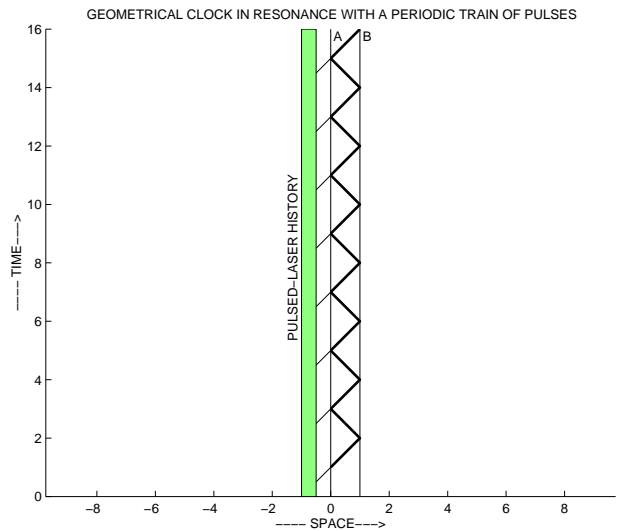}
\caption{\label{fig:free float clock} Free-float geometrical clock
  driven at resonance by a periodic train of pulses. The partial
  transmissivity of partial reflector A admits part of their amplitude
  into the interior of geometric clock AB. When their period matches
  the reflection time inside the clock, resonance prevails, and they
  form a single pulse which bounces back and forth (heavy zig-zag line)
  inside AB. This bouncing is the ticking of the geometrical
  clock. The ticking interval is the standard of time, while the
  spatial extent of AB is the associated standard of length.
Note that the picture omits, among others, the laser pulses reflected
from A and the partially transmitted pulses emerging from B.}
\end{figure}

By having such a clock control the pulse repetition rate of a
mode-locked femtosecond-laser, one generates a phase-coherent train of
pulses \cite{Udem:2002}. Introduce this train into the one-dimensional
resonance cavity of a geometrical clock with ends at relative rest
as shown in Figure~\ref{fig:free float clock}. A
resonance condition is obtained when (twice) the length of that cavity
is adjusted to equal the spacing in that train of pulses. This
resonance condition accomplishes two things:
\begin{enumerate}
\item
It establishes the relationship between a
Cesium-controlled standard of time, i.e. the duration between successive
femtosecond pulses, and the corresponding standard of length, i.e. the
size of the resonance cavity.
\item
It makes the geometrical clock into a single representative of a
standard of time \emph{and} of the space measurements. The periodic
tickings of the pulse bouncing back and forth inside provides copies
of that standard of time, while the adjusted cavity size furnishes
that standard of length\footnote{Geometrical clocks with with cavity
ends at relative rest ($k_{AB}=1$) where first used by R.F. Marzke and
J.A. Wheeler \cite{Marzke_and_Wheeler} and advocated by them as a standard of length.}.
\end{enumerate}

\section{COMMENSURABILITY}

Recall that a geometrical clock is defined as a pair of radar units
whose emission of monochromatic wave trains yields two well defined
mutual frequency shift factors ($k_{AB}$ at B and $k_{BA}$ at A) and whose
reflective surfaces support an electromagnetic pulse bouncing back and
forth between them.

Geometrical clocks differ from one another by virtue of their
differing frequency shift factors and hence their differing ticking
rates.  Moreover, these rates are in general not even
uniform. Nevertheless a comparison of these clocks is possible. The
general idea is to consider the \emph{ratio} of their ticking rates.
These ratios open the door to identifying a commensurable property
even among certain geometrical clocks which run nonuniformly relative
to their resident atomic clocks. The implementation of this endeavor
is best done in two steps: first for adjacent clocks, then for distant
ones.

\subsection{Adjacent Clocks}

To compare the operation of two \emph{adjacent} geometrical clocks,
AB and BC one notes that they have radar unit B in common. Assume that all three
radar units A, B, and C move collinearly along the $z$ axis. The
common radar unit B has electromagnetic pulses bouncing off it.
There are those from A and those from C. Consider $a$
consecutive pulses from A and $c$ consecutive pulses from C:
\begin{widetext}
\begin{eqnarray*}
&\begin{array}{c}
~\\
\textrm{pulses at B coming from A:}
~~~~
\end{array}&
\begin{array}{c}
~\\
\bullet
\end{array}
\cdots
~
\begin{array}{c}
p_i\\
\bullet
\end{array}
\overbrace{
\begin{array}{c}
p_{i+1}\\
\bullet
\end{array}
\begin{array}{c}
p_{i+2}\\
\bullet
\end{array}
\begin{array}{c}
p_{i+3}\\
\bullet
\end{array}
~
\cdots
~
\begin{array}{c}
p_{i+a}\\
\bullet
\end{array}
}^{a}
~
\begin{array}{c}
p_{i+a+1}\\
\bullet
\end{array}
~
\cdots
~
\begin{array}{c}
~\\
\bullet
\end{array}
\cdots\\
&~&~~~~~~\rule[.8mm]{3.5cm}{.16mm}~
\textrm{(time)}~\rule[.8mm]{3.5cm}{.16mm}\!\!\!\!\!\!\rightarrow ~\\
&\begin{array}{c}
~\\
\textrm{pulses at B coming from C:}
~~~~
\end{array}&
\begin{array}{c}
~\\
\bullet
\end{array}
\cdots
\,
\begin{array}{c}
~
q_j\\
\bullet
\end{array}
~
\underbrace{
\begin{array}{c}
~
q_{j+1}\\
\bullet
\end{array}
~~~
\begin{array}{c}
q_{j+2}\\
\bullet
\end{array}
\begin{array}{c}
~\\
~
\end{array}
\cdots
~
\begin{array}{c}
q_{j+c}\\
\bullet
\end{array}
}_{c}
\begin{array}{c}
q_{j+c+1}\\
\bullet
\end{array}
~
\cdots
~
\begin{array}{c}
~\\
\bullet
\end{array}
\cdots~.
\end{eqnarray*}
\end{widetext}
These sequences are depicted in Figures \ref{fig:two inertial clocks} and
also in \ref{fig:two accelerated clocks}. We say that these two sequences
are \emph{matched relative to} B, and we write
\[
\{p_i,p_{i+1},\cdots ,p_{i+a}\}_B \sim \{q_j,q_{j+1},\cdots , q_{j+c}\}_B~,
\]
if and only if they have -- within a prespecified accuracy --
coincident starting ($p_{i}$ and $q_{j}$) and coincident termination
($p_{i+a}$ and $q_{j+c}$) pulses. The subscript B on these sequences
serves as a reminder that the pulses are being counted at radar unit
B. 

The electromagnetic pulses impinging on B get partially reflected and
partially transmitted. Thus for every pulse sequence
$\{p_i,p_{i+1},\cdots ,p_{i+a}\}_B$ measured at B there are
corresponding sequences $\{p_i,p_{i+1},\cdots ,p_{i+a}\}_A$ and
$\{p_i,p_{i+1},\cdots ,p_{i+a}\}_C$ measured at A and C respectively.
Thus one has the following proposition (``Invariance of matched sequences''):

\noindent
\emph{The property of being matched is invariant as each sequence of
pulses travels from one radar unit to another,} i.e. if
\[
\{p_i,p_{i+1},\cdots ,p_{i+a}\}_B \sim \{q_j,q_{j+1},\cdots , q_{j+c}\}_B~,
\]
then
\[
\{p_i,p_{i+1},\cdots ,p_{i+a}\}_A \sim \{q_j,q_{j+1},\cdots , q_{j+c}\}_A
\]
and 
\[
\{p_i,p_{i+1},\cdots ,p_{i+a}\}_C \sim \{q_j,q_{j+1},\cdots , q_{j+c}\}_C~.
\]
The validity of this proposition is an expression of the principle of
the constancy of the speed of light, that is, of the fact that light
pulses cannot overtake each other. If two pulses, say $p_i$ and $q_j$,
are coincident on the world line of radar unit B, then they are still
coincident after they have travelled to the world line of any other
radar unit, regardless of its motion.

As measured by atomic clock B, the ticking rates of geometrical clocks
AB and BC need not be uniform, and in general they are not. This is
evident from Figure \ref{fig:two inertial clocks}.  This deficiency is
remedied by calibrating the rate of pulses coming from C in terms of
AB. Thus for every $c$-sequences of pulses departing from C and
arriving and counted at B, there is a matched $a$-sequence generated
by clock AB also at B. The ratio
\begin{equation}
\frac{c}{a}=
\frac{\textrm{(\# of ticks of clock BC)}}{\textrm{(\# of ticks of clock AB)}}
\label{eq:relative frequency}
\end{equation}
is the \emph{normalized ticking rate} of clock BC. The normalization
is relative to clock AB. 
Conversely, the inverse of Eq.(\ref{eq:relative
frequency}),
\begin{equation}
\frac{a}{c}= \frac{\textrm{(\# of ticks of clock AB)}}{\textrm{(\# of
ticks of clock BC)}}
\label{eq:inverse relative frequency}
\end{equation}
is the ticking rate of AB normalized relative to BC.  Because of the
invariance of matched sequences, it does not matter whether the
ratios, Eqs.(\ref{eq:relative frequency})-(\ref{eq:inverse relative
frequency}), were measured at radar unit A, B, or C.

We say that the adjacent clocks AB and BC are \emph{normalizable} if
both ratio (\ref{eq:relative frequency}) and ratio (\ref{eq:inverse
relative frequency}) are non-zero for every matched pair of $a$ and
$c$-sequences along the world lines of the two adjacent clocks.  A
basic and obvious aspect of normalizability for adjacent clocks is
its reciprocal property: If AB is normalizable relative to BC, then BC is
normalizable relative to AB. Thus all collinear clocks AB, BC, BD, BE,
$\cdots$, which share radar unit B, are mutually normalizable.

Of particular utility are clocks which are \emph{commensurable}.
Their distinguishing property is obtained by subdividing the set of
normalizable geometrical clocks further and selecting those whose
normalized ticking rates, Eq.(\ref{eq:relative frequency}) or
(\ref{eq:inverse relative frequency}), are \emph{constant} for all
matched starting and termination pulses.  Such clocks allow one to
view the boost-invariant accelerated and the boost-invariant expanding inertial frames
from a single perspective, which is developed in Section
\ref{sec-TRANFER OF TIME ACROSS AN EVENT HORIZON}.

Before giving the precise general definition of commensurability
(Section \ref{sec-Distant Clocks}, we interrupt the developement by
illustrating the above constellation of definitions, applying
them to various combinations of inertial and accelerated clocks.

\emph{Nota bene:} For the purpose of verbal shorthand we shall allow
ourselves to refer to ``\emph{geometrical} clocks'' simply as
``clocks''.  However, for \emph{atomic} clocks we shall use no such
shorthand.  Thus clocks without the adjective ``atomic'' are
understood to be \emph{geometrical} clocks, while \emph{atomic} clocks
are always referred to by means of the modifier ``atomic''.

\subsubsection{Commensurable Inertial Clocks}

\begin{figure}
\includegraphics[scale=.5]{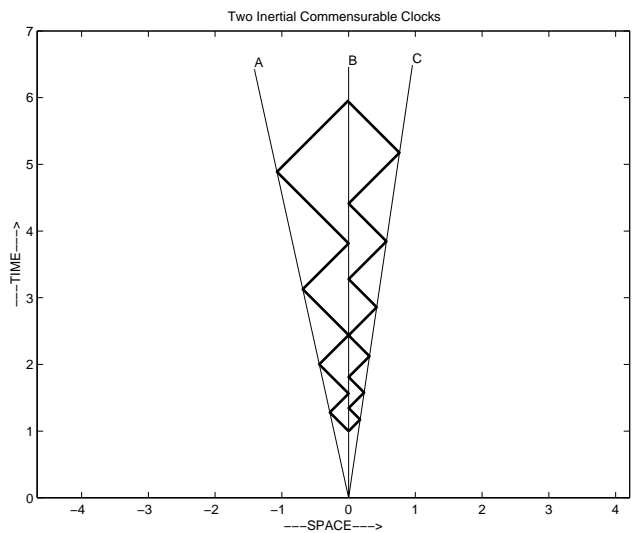}
\caption{\label{fig:two inertial clocks} Two adjacent geometrical
  clocks consisting of inertially expanding cavities AB and BC, each
  containing a pulse bouncing back and forth. Depicted in this diagram
  are sequences of 2 pulses coming from A and impinging on B which are
  matched by corresponding sequences of 3 pulses coming from C. The
  ratio $c/a=3/2$ is the ticking rate of BC normalized to that of
  AB. The fact that this ratio stays constant throughout the history
  of the two clocks makes them \emph{commensurable}.}
\end{figure}

Consider a pair of clocks AB and BC,
where all three radar units A, B, and C are freely floating, and radar
unit B is common to AB and BC, as in Fig. \ref{fig:two inertial clocks}. 

What is the ratio $c/a$ of two matched
pulse sequences impinging on radar unit B and coming from radar units
A and C?  This ratio is determined by the following mini-calculation:

Let A emit two pulses separated by 
\[
\Delta \xi
\]
as measured by atomic clock A. Due to the relative motion of A and B
these two pulses, once received at B, have time separation
\[
k_{AB}\Delta \xi
\]
as measured by atomic clock B. Here $k_{AB}$ is a positive (``Doppler'')
factor whose magnitude expresses the motion of A relative to B. There are now
two time intervals: the one between the emitted pulses and the one
between the received pulses. These intervals are proportional to the
wavelengths of emitted and received monochromatic radiation. Their
ratio,
\[
\frac{k_{AB}\Delta \xi}{\Delta \xi}=k_{AB}
\]
is the \emph{Doppler shift factor}. The two radar units are
understood to be at rest relative to each other whenever $k_{AB}=1$.
They are receding (resp. approaching) each other whenever $k_{AB}>1$
(resp. $k_{AB}<1$), which expresses a Doppler red (resp. blue) shift. It is clear
that this Doppler shift of clock AB controls the rate at which the
back-and-forth bouncing pulse produces ticks at radar unit B. In fact,
the pulse arrival times of $a$ consecutive pulses coming from A are
\[
\xi,\, k_{AB}^2 \xi, \cdots,\, k_{AB}^{2a} \xi.
\]
Similarly, the arrival times of $c$ consecutive pulses coming from
radar unit C, which is part of clock BC, are
\[
\xi,\,k_{BC}^2 \xi, \cdots, \, k_{BC}^{2c} \xi.
\]
These two pulse sequences have coincident initial pulse arrival times
$\xi$. If these two sequences are ``matched'', then their final pulse arrival times,
$k_{AB}^{2m}\xi=k_{BC}^{2n}\xi$, also coincide. Under this circumstance the
length of these two pulse sequences as measured by atomic clock B are
the same. Consequently,
\[
k_{AB}^{2a} \xi-\xi=k_{BC}^{2c} \xi-\xi,
\]
or
\begin{equation}
\frac{c}{a}= \frac{(1/\log k_{BC})}{(1/\log k_{AB})}.
\label{eq:inertial clock rate}
\end{equation}
This is the ticking rate of clock CB normalized relative to clock AB.
This ticking rate is a constant independent of the starting time $\xi$
of the two matched pulse sequences. Consequently, clock AB is
\emph{commensurable} with BC.

\subsubsection{Commensurable Accelerated Clocks}
Again consider a pair of clocks AB and BC. 
\begin{figure}[h!]
\includegraphics[scale=.5]{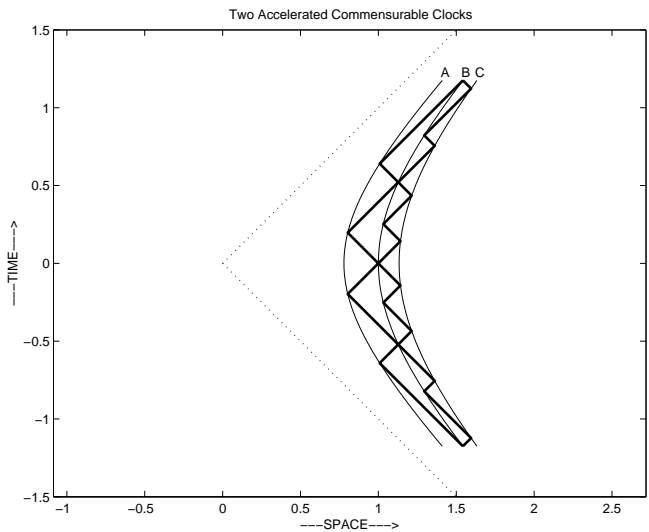}
\caption{\label{fig:two accelerated clocks} Two adjacent geometrical
  clocks consisting of collinearly accelerated cavities AB and BC,
  each containing a pulse bouncing back and forth. Depicted in this
  diagram is a sequence of 4 pulses coming from A and impinging on B
  which are matched by a sequence of 8 pulses coming from C. The ratio
  $c/a=8/4$ is the ticking rate of BC normalized to that of AB. The
  fact that this ratio stays constant throughout the history of the
  two clocks makes them also \emph{commensurable}, just like those in
  Figure \ref{fig:two inertial clocks}}
\end{figure}
But this time have all
three of their radar units accelerate collinearly to the right with
respective constant accelerations $1/\xi_A$, $1/\xi_B$, and $1/\xi_C$
respectively, and with common future and past event horizons, as in Figure
\ref{fig:two accelerated clocks}.
To make the discussion concrete, assume that $0<\xi_A <\xi_B <\xi_C$.

Consider the ticking produced by a pulse bouncing back and forth 
in clock AB. The proper time between two successive ticks at A is
\begin{eqnarray*}
\xi_A \times \left(\begin{array}{c}
               \textrm{boost coordinate}\\
	       \textrm{time between ticks}
	     \end{array} \right)
= \xi_A \times 2\log(\xi_B/\xi_A)~,
\end{eqnarray*}
while at B it is 
\begin{eqnarray*}
\xi_B \times \left(\begin{array}{c}
               \textrm{boost coordinate}\\
	       \textrm{time between ticks}
	     \end{array} \right)
= \xi_B \times 2\log(\xi_B/\xi_A)~,
\end{eqnarray*}
Their ratio
\begin{equation}
\frac{2\,\xi_B \log(\xi_B/\xi_A)}{2\,\xi_A \log(\xi_B/\xi_A)}=
\frac{\xi_B}{\xi_A}\equiv k_{AB}
\label{eq:redshift for AB}
\end{equation}
is the \emph{pseudo-gravitational frequency shift factor}. Similarly,
for clock BC one has
\begin{equation}
\frac{\xi_C}{\xi_B}\equiv k_{BC}~.
\label{eq:redshift for BC}
\end{equation}
These two frequency shift factors control the rate at which pulses arrive at
B from A and C respectively. In fact, the two corresponding matched pulse
sequences are 
\[
0,2\,\xi_B \log(\xi_B/\xi_A),\cdots,2\,a\xi_B \log(\xi_B/\xi_A)
\]
and
\[
0,2\,\xi_B \log(\xi_C/\xi_B),\cdots,2\,c\xi_B \log(\xi_C/\xi_B)~,
\]
where the last pulse arrival time is the same, i.e.
\[
2\,a\xi_B \log(\xi_B/\xi_A)=2\,c\xi_B \log(\xi_C/\xi_B)~,
\]
or with the help of Eqs.(\ref{eq:redshift for AB}) and (\ref{eq:redshift for BC})
\begin{equation}
\frac{c}{a}= \frac{(1/\log k_{BC})}{(1/\log k_{AB})}~.
\label{eq:accelerated clock rate}
\end{equation}
This is the ticking rate of clock BC normalized relative to clock AB.
This ticking rate is a constant independent of the starting time 
of the two matched pulse sequences. Consequently, accelerated clocks
AB and BC are also \emph{commensurable}.

\subsection{Distant Clocks}
\label{sec-Distant Clocks} 
\begin{figure}
\includegraphics[scale=.5]{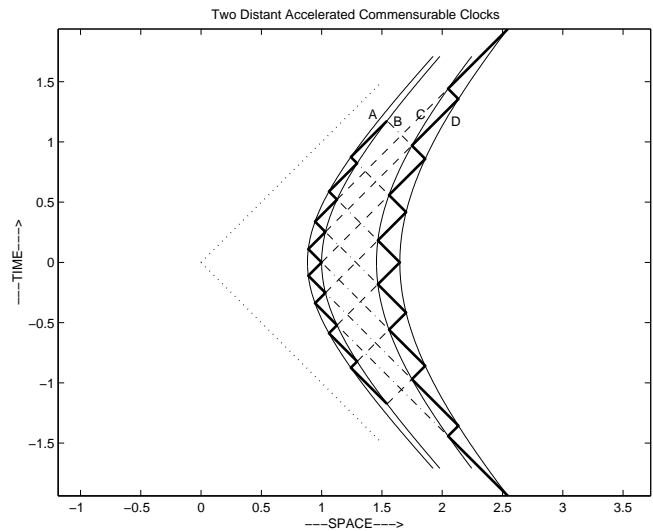}
\caption{\label{fig:two distant accelerated clocks} Two distant
  accelerated geometrical clocks. Clock CD can be calibrated in terms
  of AB because CD and AB are commensurable: Taking into account the
  ticking rate of the intermediate clock BC, one sees that the ratio
  of their rates is $\frac{1}{3}\times\frac{3}{1}=\frac{1}{1}$ in this
  figure. Because this ratio happens to be unity, these two
  commensurable clocks are said to have the additional property of
  being \emph{identically constructed}.}
\end{figure}

To compare the operation of two \emph{distant} clocks, AB and CD,
note that they have four different radar units. Assume them to be moving
collinearly along the $z$-axis such that A and D are the outer pair,
and B and C the inner pair, as in Figures \ref{fig:two distant
inertial clocks} and \ref{fig:two distant accelerated clocks}

\begin{figure}
\includegraphics[scale=.5]{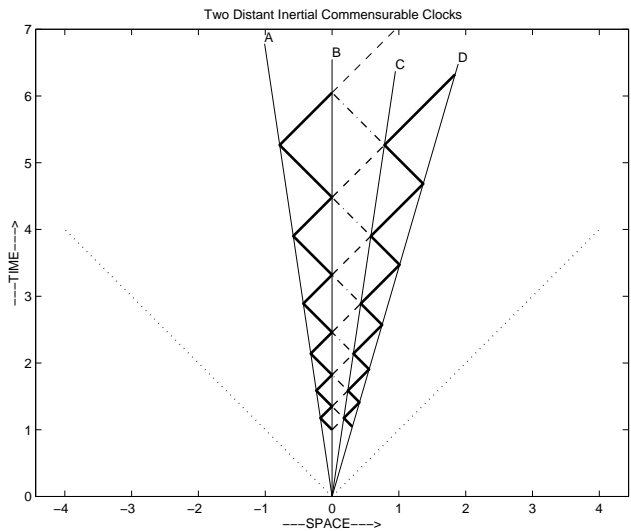}
\caption{\label{fig:two distant inertial clocks} Two distant
  inertially expanding geometrical clocks. Clock CD can be calibrated
  in terms of AB because CD and AB are commensurable: Taking into
  account the ticking rate of the intermediate clock BC, one sees that
  the ratio of their rates is a constant, namely 1:1. Because this
  ratio happens to be unity, these two commensurable clocks are said
  to have the additional property of being \emph{identically
  constructed}.}
\end{figure}

One says that two \emph{distant} (nonadjacent) clocks AB and CD are
\label{'commensurable'} \emph{commensurable}, or more briefly
\[
AB \approx CD~,
\]
if and only if 
\begin{itemize}
\item[(i)]
Radar units A and B are visible for all times to radar units C and D and
\item[(ii)]
AB is commensurable with BC, and BC is commensurable with CD. 
\end{itemize}
Being ``visible'' means that, by using its pulse radar, C can always see B
on its radar screen, i.e. BC forms a geometrical clock.
Thus two clocks AB and CD are commensurable if the clock formed by
radar units B and C is commensurable with both of its neighbors, AB and
CD.

According to this definition, one uses the constancy of the rate of clock CD
normalized to that of clock BC,
\[
\frac{(1/\log k_{CD})}{(1/\log k_{BC})}=const~,
\]
and the constancy of the rate of clock BC
normalized to that of clock AB,
\[
\frac{(1/\log k_{BC})}{(1/\log k_{AB})}=const~,
\]
to establish that the rate of clock CD
normalized to that of clock AB,
\begin{equation}
\frac{(1/\log k_{CD})}{(1/\log k_{AB})}=const~,
\label{normalized ticking rate of CD}
\end{equation}
is also a constant, and therefore that CD is commensurable with AB.
One sees from Eqs.(\ref{eq:inertial clock rate}) and
(\ref{eq:accelerated clock rate}) that this criterion for
commensurability holds for both inertial and accelerated clocks, as is
depicted in Figures \ref{fig:two distant inertial clocks} and
\ref{fig:two distant accelerated clocks}.

Commensurability of distant clocks subsumes that of adjacent clocks as
a special case. This follows from simply letting the space between
clocks AB and CD in Figures \ref{fig:two distant inertial clocks} and
\ref{fig:two distant accelerated clocks} shrink to zero so that the
final result is two adjacent clocks as in Figures \ref{fig:two
inertial clocks} and \ref{fig:two accelerated clocks}. The
commensurability is readily preserved throughout this limiting
process.

Commensurability is a relation which satisfies the following three
properties\footnote{These three properties, reflexivity, symmetry, and
transitivity, make this relation what in mathematics is called an
equivalence relation. It divides the set of clocks into mutually
exclusive equivalence classes. In our context these classes are the
various boost-invariant sectors, whose clocks can be synchronized in each
sector.  This synchronization is highlighted in Sections
\ref{sec-common method} and \ref{sec-equivalence with radar}.}:
\begin{enumerate}
\item
AB $\approx$ AB
\item
AB $\approx$ CD implies CD $\approx$ AB
\item
AB $\approx$ CD together with CD $\approx$ EF implies AB $\approx$ EF
\end{enumerate}
A physicist can choose one of these commensurable clocks as his
primary standard.  It is a dual function device : It represents a
temporal standard and a spatial standard at the same time. The
spatial extent of the clock is determined uniquely by its ticking
rate, a light pulse bouncing back and forth between the clock's two
ends.

\section{MEASURING EVENTS VIA RADAR}
Assume the physicist has chosen a standard clock whose cavity ends have
relative frequency shift factor
\[
k_{AB}\equiv e^{\Delta\tau}~.
\]
This Fourier compatibility factor controls the clock rate, which in
turn controls the times that outgoing pulses leave the clock and the
times that the receiver is turned on to allow the reception and
recording of incoming pulses.

Events have a commensurable property, a property which is reducible
(by a process of counting) to a standard of length and of time. The
\emph{common method} of measuring events, and historically the first,
relies on counting replicas of a standard of length and synchronized
replicas of a standard of time in order to identify an event in terms
of coordinates. The standard of length and of time were considered
distinct and independent of each other. But Einstein, by a process of
hard work, noticed that (i) it takes an act of using ones
\emph{visual} faculties, and hence a familiarity with the properties
of light, and that (ii) one must have a clear understanding of what is
meant by ``looking at clocks which are synchronized'', before one can
claim to have measured an event\footnote{See, for example, chapter 7
in \cite{Einstein:Relativity} for ``a clear explanation that anyone
can understand''.}.

With that observation it became clear that the standard of length and
the standard of time cannot be chosen independently, but are related
by the speed of light.

Suppose radar had been invented before 1905. Then using the
\emph{method of radar} to measure events would have forced a physicist
to confront and solve the problem of using synchronized clocks before
1905. He would have immediately found from his observations that the
standard of length is related to the standard of time. Further more,
by following Bondi \cite{Bondi}, the physicist would have immediately
formulated the kinematics of special relativity, saving himself the
hard work that Einstein had to do.

Both the method of radar and the common method for measuring an event
have been formulated for radar sets, clocks, and measuring rods which
are unaccelerated and static relative to one another. Can one extend
these two formulations and will they remain equivalent if on drops
these restrictions? The following sections give an affirmative answer.

\subsection{The Radar Method} 

Let $n_1$ be the integer that labels a pulse emitted by
radar unit B. If that pulse gets reflected, or partially reflected,
by a scattering event, then let $n_2$ be the integer which the clock
assigns to the reflected pulse as it enters the radar receiver.
These two integers,
\begin{equation}
(n_1,n_2)~,
\label{eq:radar data}
\end{equation} 
generated by this radar ranging process are the radar coordinates of
the scattering event. They assign a unique
spacetime location to the scattering event, namely $(t,z)$ as determined by 
\begin{equation}
\begin{array}{cc}
     T_1-Z_1=t-z&\textrm{(pulse moving to the right)}\\
     T_2+Z_2=t+z&\textrm{(pulse moving to the left)}
\end{array}
\label{eq:synchronization condition}
\end{equation}
Here $(T_1,Z_1)$ and $(T_2,Z_2)$ are those two events at radar unit B
which mark the emission and the reception of a pulse at B. 

\subsubsection{Inertially Expanding Clock}

If B is controlled
by the ticking of an \emph{inertially expanding} clock, then these events are
\[
(T,Z)=be^{n\Delta\tau} (1,0), ~~~~~~~n=n_1,n_2~.
\]
These two events are marked by a square and a diamond in Figure
\ref{fig:array of expanding clocks}. The constant $b$ is the proper time 
corresponding to $n=0$.

It follows from Eqs.(\ref{eq:synchronization condition}) that the scattering
event measured by B is
\begin{eqnarray}
\lefteqn{(t,z)=}\label{eq:event in F}\\
&&be^{\Delta\tau(n_1+n_2)/2}\left(\cosh \frac{n_2-n_1}{2}\Delta\tau ~,~
\sinh \frac{n_2-n_1}{2}\Delta\tau\right) \nonumber
\end{eqnarray}
These three events, $(T_1,Z_1)$, $(t,z)$ and $(T_2,Z_2)$, are marked in 
Figure \ref{fig:array of expanding clocks} by a square, a circle, and a 
diamond.

\subsubsection{Accelerated Clock}

Similarly, if B is controlled by an \emph{accelerated} clock, then the 
corresponding three events are 
\[
(T,Z)=b(\sinh{n\Delta\tau},\cosh{n\Delta\tau}), ~~~~~~~n=n_1,n_2~.
\]
and
\begin{eqnarray}
\lefteqn{(t,z)=}\label{eq:event in I}\\
&&be^{\Delta\tau(n_1-n_2)/2}\left(\sinh \frac{n_2+n_1}{2}\Delta\tau ~,~
\cosh \frac{n_2+n_1}{2}\Delta\tau\right).\nonumber
\end{eqnarray}
They are marked in an analogous way in Figure \ref{fig:array of accelerated clocks}.

Equation (\ref{eq:event in F}) or (\ref{eq:event in I}) relates in
mathematical form a scattering event to the ticking of a single
geometrical clock chosen as a standard. This relationship has been
established by the method of radar in terms of the integers $n_1$ and
$n_2$.

\subsection{The Common (Non-Radar) Method}
\label{sec-common method}

There is, of course, the more common and familiar method. It does not
use radar. Instead, it  uses two distinct standards, namely,
identically constructed clocks and standard
rods \cite{Taylor_and_Wheeler:1992}. The measuring procedure itself, we
recall, consists of (i) locating the event by counting standard rods,
and (ii) determining its time by counting at that location the ticks
of the clock, which is synchronized to the standard clock.

One is now confronted with a question of consistency: Is this common
non-radar method compatible with the radar method, even if the spacetime
framework is based on inertially expanding or accelerated clocks?

Consider the common method of measuring an event. It consists of
starting with a geometrical clock having a spacetime history as depicted
in Figure \ref{fig:inertial clock plus nullrays} or
\ref{fig:accelerated clock plus nullrays}.  Such a clock is a standard
of time \emph{and} of length. Thus a physicist forms a spatial array of
adjacent clocks AB, BC,$\cdots$, EF,$\cdots$ which are
\emph{identically constructed} and synchronized. The definition is as follows:
\begin{itemize}
\item[~]
Clocks AB, XY, $\cdots$ are said to be \emph{identically constructed} if
their frequency shift factors are all the same: 
\begin{equation}
k_{AB}=k_{XY}=\cdots\equiv e^{\Delta\tau}~.
\label{eq:equal frequency factors}
\end{equation}
\end{itemize}
This definition is illustrated in Figures \ref{fig:two distant inertial clocks} 
and \ref{fig:two distant accelerated clocks}.

The ticking of adjacent clocks is synchronized by synchronizing the
pulses impinging on their shared radar unit. This procedure guarantees
synchronization of all clocks. It is exemplified in Figure
\ref{fig:two distant inertial clocks}. There the three clocks AB, BC,
and CD have the phases of their internal pulses adjusted to tick in
synchrony.

Suppose standard clock AB has its $n$th (resp. $(n+1)$st) ticking event
at its left (resp. right) radar unit A (resp. B). These events are
exhibited by Eq.(\ref{eq:inertially expanding standard}) or
(\ref{eq:accelerated standard}).  Then, by induction, the left radar
unit of the $M$th identically constructed clock has its $N$th ticking
event at
\begin{widetext}
\begin{eqnarray}
(t,z)&=&
be^{N\Delta\tau}\left(\cosh M\Delta\tau ~,~\sinh M\Delta\tau\right)~~~~
     (\textrm{for the }M\textrm{th inertially expanding clock})
     \label{eq:inertially expanding clock event}\\
(t,z)&=&
be^{M\Delta\tau}\left(\sinh N\Delta\tau ~,~\cosh N\Delta\tau\right)~~~~
     (\textrm{for the }M\textrm{th accelerated clock}) 
     \label{eq:accelerated clock event}
\end{eqnarray}
\end{widetext}
Having formed a linear array of such clock, the physicist uses the
lattice of events generated by their tick-tock actions as a standard
to measure an arbitrary event. The common method of measuring an event
consists of counting (i) how many clocks separate it from the standard
clock ($M=0$), and (ii) how many clock ticks elapse before this event
happens. The result of these two counts is  the pair of integers 
\begin{equation}
\begin{array}{cc}
      M=m& \left( \begin{array}{c}
	      \textrm{result of}\\
	      \textrm{spatial measurement}
	   \end{array} \right)\\
~&~\\
      N=n&\left( \begin{array}{c}
              \textrm{result of}\\
              \textrm{temporal measurement}
           \end{array} \right).
\end{array}
\label{eq:measurement result}
\end{equation}
They comprise the measurement of the given event in units of time and
spatial extent as furnished by the standard geometrical clock.

\subsection{Its Equivalence With The Radar Method}
\label{sec-equivalence with radar}

Now compare Eq.(\ref{eq:inertially expanding clock event}) with
Eqs.(\ref{eq:measurement result}) and (\ref{eq:event in F}) or
Eq.(\ref{eq:accelerated clock event}) with Eqs.(\ref{eq:measurement
result}) and (\ref{eq:event in I}). Observe that for both
cases
\footnote{If $n_2\pm n_1$ is odd, then this simply means that there
does not happen to exist a clock tick at B simultaneous with event
$(m,n)$. This is, of course, due to the fact that the clock does not
furnish half-integer ticks.}
\begin{eqnarray}
m&=&\frac{n_2 -n_1}{2}\\
n&=&\frac{n_2 +n_1}{2}~.
\end{eqnarray}
One sees that the radar method is equivalent to the common method
provided one identifies the radar pulse data $(n_2 -n_1)/2$ with the
$m$th distant clock, and $(n_2 +n_1)/2$ with its $n$th ticking event.
This equivalence is new. It extends the fundamental and familiar
result based on a lattice array of free-float clocks to (i) the case
of an array of inertially expanding clocks and to (ii) the case of a
array of accelerated clocks.  Put differently, it gives physical
validity to the concepts ``inertially expanding frame'' and
``accelerated frame''.

\begin{figure}
\includegraphics[scale=.6]{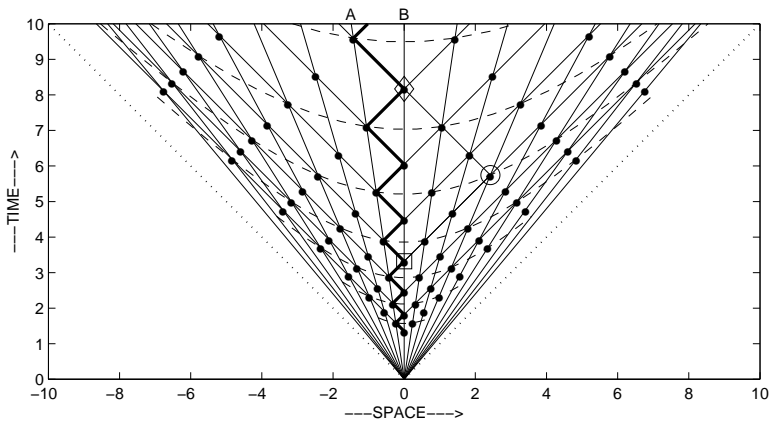}
\caption{\label{fig:array of expanding clocks}Lattice of spacetime
graduation events (heavy dots) determined and calibrated by a
single geometrical clock AB which is expanding inertially. The
spacetime history of the e.m. pulse bouncing inside this clock is the
heavy zig zag line left of the middle. The clock is bounded by two
straight lines A and B, the histories of the receding reflectors which
keep the e.m. pulse trapped inside the clock. The other straight
lines indicate the receding reflector histories of identically
constructed clocks, if they were to form an array of adjacent
geometrical clocks. The hyperbolas (dashed lines) are the times
simultaneous with the tickings of the standard clock AB.  The 45$^0$
lines emanating to the left from A and to the right from B are the
histories of the two trains of pulses escaping from A and B.  The fact
that AB is a standard clock implies that all graduation events of the
calibrated lattice lie on these histories. Based on the
method of pulsed radar, each graduation event (e.g. the encircled dot)
is labelled by two unique integers, namely two numbered ticks (the
dots in the square and in the diamond) of the clock.They are AB's
``radar coordinates'' of that graduation event.}
\end{figure}
\begin{figure}
\includegraphics[scale=.765]{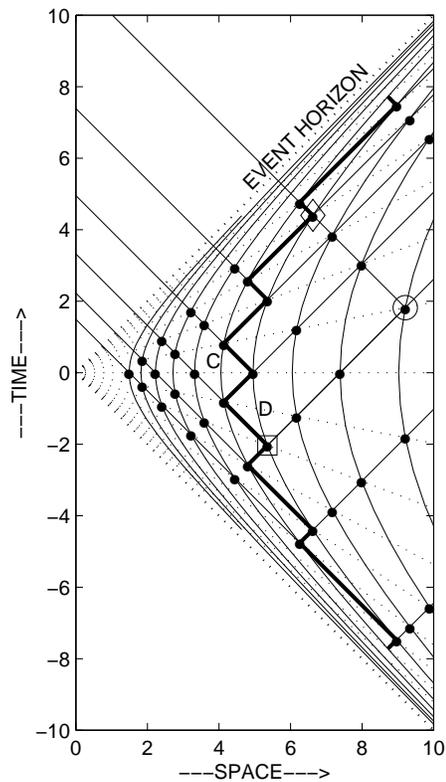}
\caption{\label{fig:array of accelerated clocks} Lattice of spacetime
graduation events (heavy dots) as determined and calibrated by a
single geometrical clock CD which is accelerating. The spacetime
history of the e.m. pulse bouncing inside this clock is the heavy zig
zag line between the two hyperbolas C and D. The clock is bounded by
these two hyperbolas, the histories of the two accelerating reflectors
which keep the e.m. pulse trapped inside the clock. The other
hyperbolas indicate the accelerated reflector histories of identically
constructed clocks, if they were to form an array of adjacent
geometrical clocks. The straight lines (lightly dotted) are the times
simultaneous with the tickings of the standard clock CD.  The 45$^0$
lines emanating to the left from C and to the right from D are the
histories of the two trains of pulses escaping from C and D, with
those escaping from C ultimately crossing the event horizon of clock
CD. The fact that CD is a standard clock implies that all graduation
events of the calibrated lattice lie on these histories. Based on the
method of pulsed radar, each graduation event (e.g. the encircled dot)
is labelled by two unique integers, namely two numbered ticks (the
dots in the square and in the diamond) of the clock. They are CD's
``radar coordinates'' of that graduation event.}
\end{figure}

\section{IDENTICALLY CONSTRUCTED CLOCKS AS SYMPATHETICALLY RESONATING CAVITIES}
\label{sec-IDENTICALLY CONSTRUCTED CLOCKS}

The existence of identically constructed clocks -- or entities which
have the properties of such clocks -- is essential for establishing a
coordinate frame. This is true not only for coordinate frames which
are freely floating, but also for those which are based on clocks
which are inertially expanding or accelerated.

One's success in realizing such clocks in the laboratory or identifying
them nature is increased considerably by the fact that their
existence is a manifestation of \emph{sympathetically resonating}
cavities.

This resonance behavior occurs when radiation is trapped in two weakly
interacting cavities with identical normal mode spectra. This behavior
plays the decisive role in the operation and synchronization of
these clocks. There are two complementary, but equivalent, ways of
understanding sympathetic resonance: in terms of travelling pulses and
in terms of normal modes.

\subsubsection{Travelling Pulses}
Consider two identically constructed clocks AB and CD, which are
characterized by the same frequency shift factors $k_{AB}$ and $k_{BA}$:
\begin{equation}
\begin{array}{c}
k_{AB}=k_{CD}\\
k_{BA}=k_{DC}
\end{array}
\label{eq:equal frequency shifts}
\end{equation}
Each of Figures \ref{fig:two distant inertial clocks} and
\ref{fig:two distant accelerated clocks} is an example of the
spacetime history of two such clocks.

Let e.m. pulses from the first ticking clock enter, by partial
transmission, the empty cavity of the second clock, which is initially
not ticking (no e.m. pulse inside). Then these entering pulses will
start a ticking process in this clock. Because of Eq.(\ref{eq:equal
frequency shifts}), this process is in perfect synchrony with the
impinging pulses. They come precisely at the right moment and have the
right phase so as to augment the interior pulse amplitude from one
tick to the next
\footnote{For a cavity with ends at relative rest, this is, in fact,
what happens in a Ti-doped sapphire laser.  When turned on, usually
only one or two modes are excited. Consequently, it starts its
operation in in a continuous wave mode. However, by shining light
pulses into this laser, the lasing action starts in other cavity
modes. Since Ti-doped sapphire is a broadband amplifying medium, it is
capable of sustaining this lasing action. The superposition of these
lasing modes constitutes a light pulse bouncing back and forth inside
the cavity. This bouncing is in perfect synchrony with the external
light pulses that have been shined into the laser.}
. As a result, CD starts ticking in sympathy with AB. 

If there is no event horizon between the two clocks, then the process
can be reversed, and AB ticks in sympathy with CD. When both processes
happen simultaneously, we say that the sympathetic resonance of AB and
CD is mutual. In that case the pulses carry information both
ways. This allows the synchronization of the two identically
constructed clocks.

\subsubsection{Normal Modes} 

The complementary, but equivalent (via Fourier synthesis), perspective
on this resonance is to note that the two clocks have identical normal
mode spectra.  More explicitly, the cavities have their ends moving in
such a way that the normal modes, which are governed by the wave
equation
\[
-\frac{1}{\xi^2} \frac{\partial ^2 \psi}{\partial \tau} +
\frac{1}{\xi} \frac{\partial}{\partial \xi}\xi \frac{\partial
  \psi}{\partial \xi}=0~,
\]
vibrate (as a function of $\tau$) at the same respective rates in the
two cavities. For two accelerated clock cavities AB and CD this equality
is achieved by the condition
\begin{equation}
\ln \xi_B -\ln \xi_A=\ln \xi_D -\ln \xi_C
~~~\left(\begin{array}{c}
          \textrm{acc'd cavities in}\\
	  \textrm{spacetime sector}~ I
	   \end{array} \right),
\label{eq:accelerated boundary condition}
\end{equation}
because it yields
\[
\psi\sim e^{-i\omega_n\tau}\sin
(\omega_n\ln\xi),~~\omega_n=\frac{n\pi}{\ln\xi_B-\ln\xi_A}~.
\]

For the circumstance of two inertially expanding clock cavities AB and CD
this is achieved by the condition
\begin{equation}
\tau_B -\tau_A=\tau_D -\tau_C
~~~\left(\begin{array}{c}
          \textrm{expanding inertial}\\
	  \textrm{cavities in}\\
          \textrm{spacetime sector}~ F
           \end{array} \right),
\label{eq:expanding boundary condition}
\end{equation}
because it yields
\[
\psi\sim
\sin(\omega_n\tau)\,\xi^{i\omega_n},~~\omega_n=\frac{n\pi}{\tau_B-\tau_A}~.
\]
The first condition is precisely the conditions for clocks AB and CD
in $I$ to be identically constructed, the second one for clocks in
$F$.  Indeed, using Eq.(\ref{eq:redshift for AB}), the definition of 
$k_{AB}$, one sees that Eq.(\ref{eq:accelerated boundary condition})
reads
\begin{equation}
\ln k_{AB}=\ln k_{CD}~,
\label{eq:1st result}
\end{equation}
which coincides with Eq.(\ref{eq:equal frequency shifts}).  Similarly,
using the definition
\footnote{The Doppler shift between two bodies
A and B is given by
\[
k_{AB}=\sqrt{\frac{1+v}{1-v}}~.
\]
Here $v$ is their relative velocity, which in terms of the coordinates
of spacetime sector $F$ is given by
\[
v=\frac{\sinh(\tau_B-\tau_A)}{\cosh(\tau_B-\tau_A)}~.
\]
This yields
\[
k_{AB}=e^{(\tau_B-\tau_A)}~.
\]}
\begin{equation}
k_{AB}=e^{(\tau_B-\tau_A)}
\label{eq:Doppler shift factor}
\end{equation}
for an inertially expanding cavity in spacetime sector $F$,
one sees that Eq.(\ref{eq:expanding boundary condition})
reads
\begin{equation}
\ln k_{AB}=\ln k_{CD}~,
\label{eq:2nd result}
\end{equation}
which again coincides with Eq.(\ref{eq:equal frequency shifts}).

The results expressed by Eqs.(\ref{eq:1st result}) and (\ref{eq:2nd
result}) can therefore be summarized by the simple statement:
Identically constructed clocks are those with cavities having
identical eigenvalue spectra.
This means that the frequencies%
\footnote{For accelerated cavities one talks about \emph{temporal}
frequencies, while for inertially expanding cavities one talks about
\emph{spatial} frequencies, but frequencies nevertheless.}  
of the field oscillators in one cavity coincide with the frequencies
of those in the other.

If there is a weak mutual interaction between the cavities (i.e. the
reflectors at the cavity ends are slightly transmissive), then there
is a coupling among each pair of normal modes (field oscillators), one
in each of the two cavities. If cavity AB starts out with all the
field energy, then this coupling mediates the excitation of the
field oscillators in CD at their respective
frequencies. They will start oscillating in sympathy with those of AB.

The sum of all the (normal mode) amplitudes of these field oscillators
forms a bouncing pulse in CD. The fact that the sympathetic resonance
makes these amplitudes increase with time implies that the bouncing
pulse in CD does the same.

To summarize: An analysis in terms of bouncing pulses or in terms of
normal modes leads to the same conclusion: The physical process of the
transfer of time (a train of clock ticks) between identically
constructed clocks is the process of sympathetic resonance between
their cavities.

\section{TRANSFER OF TIME ACROSS FROM AN ACCELERATED TO AN INERTIALLY EXPANDING
  CLOCK}
\label{sec-TRANFER OF TIME ACROSS AN EVENT HORIZON}

\begin{figure}
\includegraphics[scale=.5]{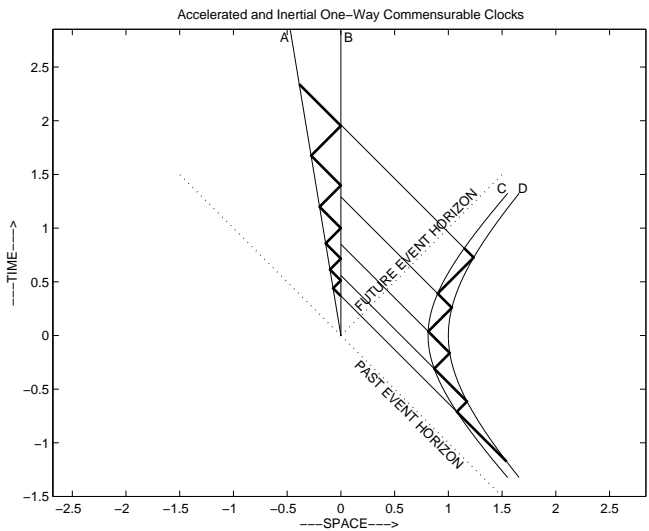}
\caption{\label{fig:accelerated and inertial clocks} One-way
  commensurability between inertially expanding clock AB and
  accelerated clock BC, each containing its pulse bouncing back and
  forth. Depicted in this diagram is a sequence of 5 high intensity
  pulses coming from A and impinging on B which are matched by a
  sequence of 4 low intensity pulses coming across the event horizon
  from C. The ratio $m/n=5/4$ is the ticking rate of AB normalized to
  that of CD. The fact that this ratio stays constant throughout the
  history of the two clocks makes them \emph{one-way commensurable}.
  }
\end{figure}

Commensurability is a more basic property than the property of clocks
being synchronized. Before one tries to synchronize two clocks,
one must first ascertain that they are commensurable.

Furthermore, two commensurable clocks cannot be synchronized unless there is a
two-way interaction between them. In the context of an inertially
expanding or an accelerated coordinate frame (Figure \ref{fig:array of
expanding clocks} or \ref{fig:array of accelerated clocks}) such an
interaction consists of a radar (to and fro) signal between each pair
of clocks, say AB and CD as in Figures \ref{fig:two distant inertial
clocks} or \ref{fig:two distant accelerated clocks}. Such a radar
signal accommodates a two-way transfer of time: AB transmits its tick
number to CD, and CD sends via the return pulse its own tick number
back to AB. With this mutual knowledge the two clocks can be
relabelled, if necessary, to give synchronized time.

However, if there is an event horizon between clocks AB and CD, then
qualitatively new considerations enter.

On one hand, at most only a one-way transfer of time is possible. The
establishment of a time synchronous to both of them is out of the
question. 

On the other hand, that event horizon brings with it a pleasant
surprise: an accelerated clock and an inertially expanding clock,
which at first sight seem to be incommensurable, are in fact
commensurable when there is an event horizon between them. In
particular one clock can (via sympathetic resonance) exert a one-way
control over the other.  Here is why:

As one can see from Figure \ref{fig:Rindler spacetime} there is an
event horizon that separates the clocks in spacetime sector $I$ from
those in spacetime sector $F$. But the problem with taking advantage of
a one-way transfer of time from CD in $I$ to AB in $F$ seems to be
that the clock in $I$ is accelerated while the one in $F$ is
inertially expanding. At first sight there seems to be no way that the
two are commensurable as defined on page~\pageref{'commensurable'} in
section \ref{sec-Distant Clocks}. One must note, however, that that
definition was based on a two-way transfer of time (``AB is
radar-visible to CD''). This was necessary. Indeed, the definition of
boost-invariant sector $I$ as well as $F$ (``equivalence classes of
geometrical clocks that can be synchronized'') depended on it.

To accommodate the context of an event horizon as a one-way membrane
between clocks AB and CD, we enlarge the concept ``commensurability''
by defining the concept ``one-way commensurability''. This is done by
dropping the requirement that radar units B and C be in two-way
contact, and by saying that one-way contact, say from C to B,
is good enough.  The result of doing this is
illustrated in Figure \ref{fig:accelerated and inertial clocks}.

Accelerated clock CD moves along the line of sight of inertially
expanding clock AB.  This clock is characterized by Doppler shift
factor $k_{AB}$.
Clock CD, whose radar units are accelerated with constant
accelerations $1/\xi_C$ and $1/\xi_D$ to the right, is
characterized by the pseudo-gravitational frequency shift factor
\[
k_{CD}=\frac{\xi_D}{\xi_C}
\]
between them. As shown in Figure~\ref{fig:accelerated and inertial
clocks}, clock CD sends pulses on a one-way journey to AB. There are
no return pulses. Nevertheless, one can compare a sequence of $a$
pulses at B from A with a matched sequence of $c$ pulses from CD.
The result is 
\begin{equation}
\frac{a}{c}= \frac{(1/\log k_{AB})}{(1/\log k_{CD})}~.
\label{eq:equal frequency shift factors}
\end{equation}
This is the ticking rate of inertial clock AB normalized relative to
accelerated clock CD.  This ticking rate is a
constant independent of the starting time of the two matched pulse
sequences. Consequently, inertial clock AB is \emph{one-way
commensurable} with accelerated clock BC.

\begin{figure}
\includegraphics[scale=.6]{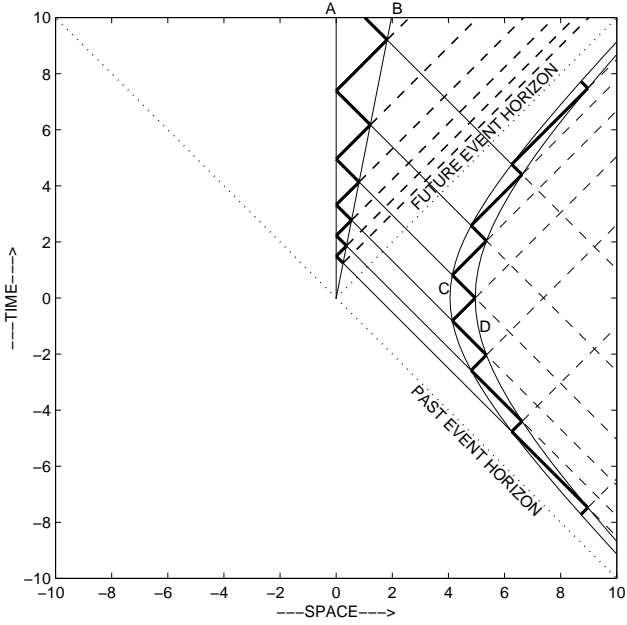}
\caption{\label{fig:resonating clocks} One-way transfer of time from
  clock CD to clock AB. This transfer is achieved by the ticking of CD
  causing sympathetic tickings in clock AB. This process is brought
  about by e.m. pulses travelling from CD across the future event
  horizon to AB (solid thin 45$^\circ$ lines).  They strike B at
  precisely the same rate and with the same phase as AB's clock pulse
  (heavy zigzag line) bouncing repeatedly off B. }
\end{figure}
Equation (\ref{eq:equal frequency shift factors}) is a remarkable
result for a number of reasons. First of all, there is its
constancy. Contrast this with the tickings of the comoving atomic clocks
at radar units B, which is floating freely, and C, which is
accelerated. They yield
\[
\frac{\textrm{(\# of ticks of atomic clock C but observed at
B)}}{\textrm{(\# of ticks of atomic clock B)}} ~,
\]
a corresponding rate which is a
Doppler chirp towards the red as seen by a physicist comoving with the
free-float atomic clock at B. By contrast, the constancy of
Eq.(\ref{eq:equal frequency shift factors}) expresses the fact that 
the slowdown in the \emph{proper}
ticking rate of geometrical clock AB compensates precisely for the
slowdown in the proper rate of pulses arriving at B from C. 

Second, if $n/m=1$, i.e.
\begin{equation}
k_{AB}=k_{CD}~,
\label{eq:same frequency shift factor}
\end{equation}
then the process of transferring a train of clock pulses from across
its future event horizon to clock AB (``one way transfer of time'') is
a process of tickings in cavity CD bringing about sympathetic tickings
in cavity AB. The implementation of this transfer is depicted in
Figure~\ref{fig:resonating clocks}.  Thus, following the discussion in
Section \ref{sec-IDENTICALLY CONSTRUCTED CLOCKS}, one concludes that,
even though CD is accelerated while AB is expanding inertially, (i)
the two cavities are \emph{identically constructed} from perspective
of their normal mode spectra, and that (ii) AB and CD \emph{are
ticking at the same rate} as measured at B.

\section{TRANSFER OF RADAR DATA ACROSS AN EVENT HORIZON}

\begin{figure}[h!]
\includegraphics[scale=.6]{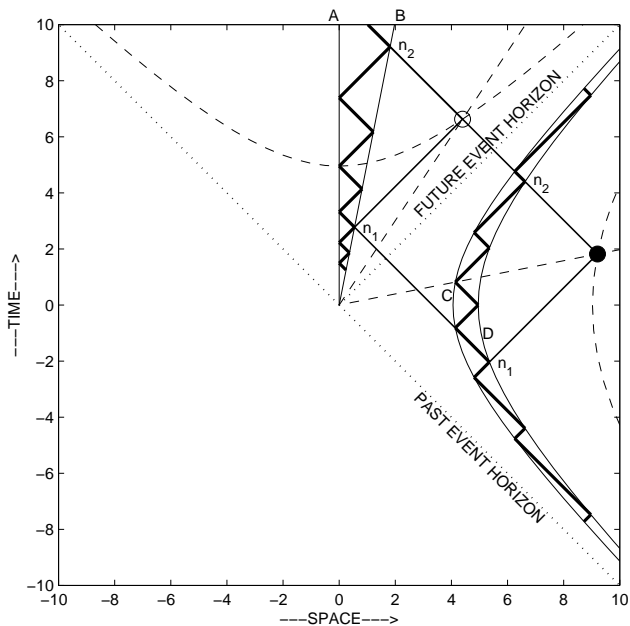}
\caption{\label{fig:radar map from I to F} Transfer of radar
  coordinates from clock CD to clock AB. The heavy dot in $I$
  gets mapped to the hollow dot in $F$. The radar coordinates of
  the event marked by the heavy black dot are the two clock pulses
  numbered $n_1$ and $n_2$ by accelerated clock CD. These two pulses
  cause CD to dispatch two pulses across its future event
  horizon. They are received by inertial clock AB where the numbering
  of its pulses is the same, namely, $n_1$ and $n_2$. These are also
  the radar coordinates of the encircled event (``hollow dot'') 
  in $F$. If this and
  other such events lie along the straight dashed line in $F$, then
  a free-float physicist who watches AB knows that the scattering 
  source in $I$
  has the world line of the dashed hyperbola in $I$.}
\end{figure}

Third, if clock CD controls the emission and reception of radar pulses
to locate a scattering event in $I$ (filled circle in
Figure~\ref{fig:radar map from I to F}), then upon being transferred
to AB, these pulses can be used by AB to reconstruct an image of that
event's location in $F$ (unfilled circle in Figure~\ref{fig:radar map
from I to F}). By applying this reconstruction to scattering events
lying on, say, a time-like hyperbola in $I$, (dashed curve in
Figure~\ref{fig:radar map from I to F}), one finds that its image in
$F$ is a straight line in $F$ (dashed line in Figure~\ref{fig:radar map
from I to F}). Similarly, a spacelike straight line of simultaneity
gets reconstructed as a spacelike hyperbola of simultaneity in F.

Mathematically this reconstruction assumes its simples form when
expressed in terms of the null coordinates
\begin{eqnarray*}
U=t-z\\
V=t+z
\end{eqnarray*}
of Figure~\ref{fig:Rindler spacetime}:
\begin{widetext}
\begin{eqnarray*}
I&\longrightarrow& F\\
\left(
 \begin{array}{c}
    \textrm{Scattering}\\
    \textrm{event in }I
 \end{array}
\right):~(U_I,V_I) &\sim\!\rightarrow& \left(
                                             \begin{array}{c}
					       \textrm{Image of}\\
					       \textrm{scattering}\\
					       \textrm{event in }F
					     \end {array}
				       \right):~(U_F,V_F)=\left( \frac{b^2}{-U_I}, V_I \right)\\
(\textrm{hyperbola in }I):~U_IV_I=const.&\longrightarrow& \left(
 \begin{array}{c}
    \textrm{Straight}\\
    \textrm{line in }F
 \end{array}
\right):~\frac{V_F}{U_F}=\frac{-const.}{b^2}\\
\left(
 \begin{array}{c}
    \textrm{Straight}\\
    \textrm{line in }I
 \end{array}
\right):~\frac{V_I}{U_I}=const. &\longrightarrow& (\textrm{hyperbola
in }F):~U_FV_F=-b^2\times const.
\end{eqnarray*}
\end{widetext}

Physically this reconstruction is based entirely on $1/b$, the
acceleration of radar unit C of clock CD, and Eq.(\ref{eq:same
frequency shift factor}), the common frequency shift factor, which is
best expressed in terms of the change in the boost coordinate $\tau$,
\[
e^{\Delta\tau}\equiv k_{AB}=k_{CD}~.
\]
Let 
\begin{eqnarray*}
n_2=n+m\\
n_1=n-m
\end{eqnarray*}
be the integer-valued radar coordinate of a scattering event located
by CD in $I$. Then that event is related to its image in $F$ by
\begin{widetext}
\begin{eqnarray*}
I&\longrightarrow& F\\
(t_I,z_I)=be^{m\Delta\tau}(\sinh n\Delta\tau ,\cosh n\Delta\tau)&\sim\!\rightarrow&
(t_F,z_F)=be^{n\Delta\tau}(\cosh m\Delta\tau ,\sinh m\Delta\tau)
\end{eqnarray*}
\end{widetext}
This is the relationship between the solid and the hollow circled
events in Figure \ref{fig:radar map from I to F}.  

\section{SUMMARY}

The subject of this article is inertially expanding reference frames.
The theme is their establishment as a one-way windows into uniformly and
linearly accelerated reference frames. To execute this task, this
article constructs both kinds of frames using the basic properties of
Doppler radar and pulse radar and then points out a radar-generated
mapping between the two frames.

The construction requires three ingredients.  First one constructs
``Fourier-compatible'' geometrical clocks.  Each one is characterized
by a single number, the frequency shift factor between the moving ends
of a one-dimensional moving cavity that traps an e.m. pulse. Its
bouncing action provides the ticking of the clock.

Second one introduces geometrical clocks. In general, the ticking
rates of these clocks are non-uniform relative to an atomic clock
comoving with one end of the cavity, or the other. Thus the necessity
of comparing one geometrical clock with another is fulfilled by
considering the \emph{ratio} of theses rates. This leads to the
concept ``commensurability'' as applied to geometrical clocks.  They
are commensurable whenever one can choose one of them as a
representative with its pair of properties (separation between
successive ticks and separation between cavity ends) serving as a
standard: the properties of all other clocks can be related to it
numerically. Thus ``commensurability'' is a relation which implies and
is implied by a standard of measurement. Moreover,
``commensurability'' is an equivalence relation. It divides clocks
into mutually exclusive sets, which in mathematics are called
equivalence classes and in physics are called reference frames,
accelerated ($I$ or $II$ as in Figure~\ref{fig:Rindler spacetime}) or
inertially expanding ($F$ or $P$ as in Figure~\ref{fig:Rindler
spacetime}).

It is valuable to take note of the importance 
of commensurability as a general concept and as a special concept when
applied to geometrical clocks in particular. It is the connecting link
between nature (i.e. reality, existence) and the observer's mind. This
is because the observer can choose one of these clocks as a standard
to which he can quantitatively relate all others (by taking ratios).
Then, by referring to merely a single representative clock the
observer can grasp the corresponding equivalence class of all possible
commensurable clocks, a particular spacetime coordinate frame
(accelerated or inertially expanding).

Without a measurement process of some sort, there would not be a
commensurability criterion, there would not be an equivalence
relation, hence no equivalence class, i.e. no concept.  The concept of
time and of place consists of the set of measurement results which the
observer obtains when he relates the tickings and the size of his
standard clock to any event occurring in his reference
frame. Establishing these relations is what he means by measuring (the
time and place of) these events.

The third ingredient consists of specifying some sort of measurement
process. Just as a one-dimensional array of calibrated graduation
marks on a measuring rod facilitates measuring the length of any
specific object, so a lattice of calibrated graduation events in a
spacetime coordinate frame facilitates measuring the time and position
of any specific event.  The calibration process can be performed by
the radar method, which is based on having a single standard clock
control the emission and the reception of radar pulses, or by the
common (non-radar) method, which is based on counting synchronized
identically constructed clocks (i.e. copies of the clock chosen as a
standard) and their ticks. Even though the equivalence of these two
methods extends to accelerated as well as inertially exspanding clocks,
the introduction of radar does not make identically constructed clocks
obsolete or useless.

On the contrary. Suppose one applies the essential aspect of being
``identically constructed'' to two clocks separated by an event
horizon between them. ``Identical construction'' means that, even
though one clock is accelerated while the other is inertially
expanding, their cavities have identical eigenfrequency spectra, and
that, as a consequence, the timing pulses emitted by the accelerated
clock cause the inertially expanding clock to tick in perfect
synchrony with their arrival at this clock.

Thus the first useful aspect of two ``identically constructed'' clocks,
one accelerated the other inertially expanding, is that they lend
themselves to being (one-way) synchronized even though they are
separated by an event horizon.

The second and more important aspect is that a physicist in the
inertially expanding frame can ``look'' into the accelerated frame on
the other side of the event horizon and ``see'' the spacetime
trajectories of sources in that frame. This is because the two clocks
serve to (one-way) transfer radar images across the event horizon. The
elements (pixels) of a radar image are in the form of the amplitudes
of the pulses reflected by a scatterer located in the frame of the
accelerated clock. Taking advantage of its transponder capability, the
accelerated clock forwards these pulses to the identically constructed
inertially expanding clock. There the pulses are used to reconstruct a
spacetime image of what the accelerated clock sees. For example,
suppose the radar controlled by this clock measures that a localized
scatterer has the history of a hyperbolic world line in
boost-invariant sector $I$. Once the pixels of this radar image have
been sent across the event horizon, the inertially expanding clock
reconstructs them into a straight timelike line in boost-invariant
sector $F$. As a second example, the pixels of a linear array of
simultaneous scattering events in $I$, upon transmission across the
event horizon, get reconstructed as a spacelike hyperbola in $F$.

Thus, by using an accelerated and an inertially
expanding clock which are ``identically constructed'', an inertial
observer can verify by radar whether the dipole source in the
augmented Larmor formula is accelerated in a uniform and linear way.

\section{THREE CONCLUSIONS}
The augmented Larmor formula, Eq.(\ref{eq:geometrical Larmor
formula}), is a mathematical relation between a dipole source located
in an accelerated frame and an integral of the Poynting vector
observed in the inertially expanding reference frame, a
relation between cause and effect.

The Poynting integral is a quantity quadratic in the e.m. field
$\{\vec E (x),\vec B(x) \}$. Recall that to determine this quantity
experimentally and to validate it as a Maxwell field requires \emph{two}
distinct measuring processes. The first one measures the magnitude and
the direction of the e.m. field quantities.  This is usually done with
an antenna, a radio receiver, and a volt meter. The second one
ascertains the place and the time of this receiving antenna in
relation to the dipole source. This is usually done optically. The
physicist illuminates his receiving antenna with optical radiation.

The experimental determination of the e.m. field quantities consists
then of establishing a quantitative relation between the results of
the two measuring processes, the optical
measurements and those obtained with the receiving antenna.  The
augmented Larmor formula, Eq.(\ref{eq:geometrical Larmor formula}), is
an example of such a relation.  The independent variable $\tau$ is
measured optically. The dependent variable (flow of radiant energy
into the direction of acceleration) is measured with the receiving
antenna.  The $\tau$-measurements consist of identifying the
relationship between observation events in $F$ and the tickings of the
expanding reference clock AB in Figure~\ref{fig:array of expanding
clocks}. This clock also serves to make the $\tau$-measurements of the
source events in $I$, but only after their coordinates have been
transferred (by means of the ``radar map'') from $I$ to $F$ as
illustrated in Figure~\ref{fig:radar map from I to F}.

An obvious feature of Eq.(\ref{eq:geometrical Larmor formula}) is that
it differs from the standard Larmor formula by a significant
contribution. However, one should \emph{not} conclude from this that there is
any contradiction with established knowledge.  This is because the
prominent assumption that went into the derivation of
Eq.(\ref{eq:geometrical Larmor formula}) is that the measurement of
the e.m. Poynting integral is done in an \emph{inertially expanding
coordinate} frame. By contrast, the standard Larmor radiation formula
assumes that the measurement of the e.m. Poynting vector is done in a
free-float (``inertial'', ``Lorentz'') coordinate frame.

The difference between the standard and the augmented Larmor formula
goes with the difference between a free-float and an inertially
expanding reference frame.  These two frames are
\emph{incommensurable}. An attempt to evade this difference by, for
example, invoking a coordinate transformation between Minkowski and
boost coordinates or by resorting to ``covariance'' would be like
trying to transform apples into oranges.  The two coordinate frames
reveal entirely different aspects of nature, and the radiation from an
e.m. source is one of them.

\subsection{Boost Coordinates as Physical and Nonarbitrary}

What one \emph{should} conclude is the opposite of what some authors
have asserted in the past. For example, they claimed that ``$\cdots$
the coordinates that we use [for computation] are arbitrary and have
no physical meaning''\footnote{Remark by E. Wigner on page 285 in the
discussion following papers by S.S. Chern and T. Regge in
\cite{Wolf1980}} or ``It is the very gist of relativity that anybody
may use any frame [in his computations].''\footnote{Page 20 in
\cite{Schroedinger1956}} Without delving into the logical
fallacies\footnote{One of them, the fallacy of the ``stolen concept'',
deserves special mention because of its ubiquity, even among
physicists. It is exemplified in statements such as (i) ``coordinates
are unphysical'', (ii) ``before $t=0$ the universe did not exist''
(iii) ``the beginning of the universe'', (iv) ``the creation of the
universe'', (v) ``the birth of the universe'', (vi) ``Why does the
universe exist?'', etc.}  underlying these claims, one should be aware
of their unfortunate consequences. They tend to discourage attempts to
understand natural processes whose very existence and identity one
learns through measurements and computations based on nonarbitrary
coordinate frames.  The identification of radiation from bodies with
extreme acceleration is a case in point. For these, two complementary
frames are necessary: an accelerated frame to accommodate the source
(boost-invariant sector $I$ and/or $II$) and the corresponding
expanding inertial frame (boost-invariant sector $F$) to observe the
information carried by the radiation coming from this source. These
frames are physically and geometrically distinct from static inertial
frames. They also provide the logical connecting link between the
concepts and the perceptual manifestations (via measurements) of these
radiation processes. Without these frames the concepts would not be
concepts but mere floating abstractions.

\subsection{Conjoint Boost-Invariant Frames as an Arena for Scattering Processes}

The most prominent feature of radiation from a body with extreme
acceleration is the kinematics necessary for its observation. One
needs \emph{two} coordinate frames: one for the accelerated source, the
other for the inertial observer. It is vital that these two frames be
aligned properly (as in Figure~\ref{fig:resonating clocks}) so that
the geometrical clock of one frame is (one-way) commensurable with the
clock of the other. This commensurability locks the two frames into a
conjoint coordinate frame with an event horizon between them.

This conjointness opens vistas which are closed to the familiar
inertial frames with their respective lattice work of free-float
clocks and rods \cite{Taylor_and_Wheeler:1992}. An obvious example is
the measurement of the acceleration radiation scattered by a dipole
oscillator as it accelerates through the e.m. field in its Minkowski
vacuum state. The augmented Larmor formula applied to this oscillator
yields the result that it scatters black body radiation with 100\%
spectral fidelity relative to the inertially expanding reference
frame.

\subsection{Boost Coordinate Frame as a Valid Coordinate Frame in Quantum Field Theory} 

The consistent use of geometrical clocks puts constraints on the
mathematical formulation of waves propagating in the inertially
expanding coordinate frame $F$. In this frame, a standard inertially
expanding clock AB characterized by Doppler frequency shift factor
Eq.(\ref{eq:Doppler shift factor}),
\[
k_{AB}=e^{(\tau_B-\tau_A)}\equiv e^{\Delta\tau},
\]
generates pulses whose null histories as depicted in
Figure~\ref{fig:inertial clock plus nullrays} are 
\begin{eqnarray}
\xi e^{\tau}&=&be^{n_2\Delta\tau}~~~~~n_2=0,\pm 1, \cdots~~\\
\xi e^{-\tau}&=&be^{n_1\Delta\tau}~~~~~n_1=0,\pm 1, \cdots~~.
\end{eqnarray}
The graduation events calibrated by this geometrical clock 
yield therefore the following discrete boost coordinates
\begin{eqnarray}
\xi&=&be^{N\Delta\tau}, ~~~~~N=\frac{n_2+n_1}{2}
\label{eq:time of sampling events}\\
e^\tau&=&e^{M\Delta\tau}, ~~~~~M=\frac{n_2-n_1}{2}~.
\label{eq:location of sampling events}
\end{eqnarray}
As illustrated in Figure~\ref{fig:array of expanding clocks} and
discussed in Section \ref{sec-common method}, they are the boost
coordinates of the $M$th identically constructed clock with its $N$th
ticking event.

In a paper some time ago \cite{Padmanabhan:1990} Padmanabhan
considered the evolution of normal modes of the wave equation $(\Box
-m^2)\psi=0$ in the boot-invariant coordinate frame $F$. 

Starting with a normal mode characterized by positive boost
frequency in the distant past of $F$, he observed that this mode, in
compliance with the wave equation, evolved into a mixture of positive
and negative frequencies in the distant future of $F$. From the
viewpoint of quantum theory such a mixture indicates a production of
particles and antiparticles.  This formulation of waves propagating in
$F$ therefore leads to the mathematical prediction that, in analogy
with Parker's particle-antiparticle creation mechanism
\cite{Parker:1982} due to a time-dependent gravitational field,
particles and antiparticles get created because of the time-dependence
of the boost-invariant metric, Eq.(\ref{eq:Rindler F metric}), in $F$.

This prediction is, of course, invalid. It contradicts the
absence of any such particle creation in flat spacetime, where there
is no gravitational field. But the procedure leading to this
contradiction, Padmanbhan points out, is mathematically sound and
\emph{completely conventional} [our emphasis]. In order to avoid
this contradiction he proposes that, within the context of quantum
theory (i.e. particle-antiparticle production), one exclude Bondi and Rindler's
spacetime coordinatization as physically inadmissible.

However happy one must be about the scrutiny to which that
coordinatization has been subjected, one must not forget that
Padmanabhan's procedure leading to to the above contradiction is far
from ``completely conventional''. In fact, it violates the central
principle of measurement (Section \ref{sec-Principles of
Measurement}): ``once a standard of time has been chosen, it becomes
immutable for all subsequent measurements''. Here is how the violation
occurs:

In spacetime sector $F$, where the invariant interval has the form
\begin{equation}
ds^2=-d\xi^2+\xi^2d\tau^2+dy^2+dx^2~,
\label{eq:Rindler F metric}
\end{equation}
the normal modes of the wave
equation $(\Box-m^2)\psi=0$ have the form
\[
\psi=\psi_k(\xi,\tau)e^{i(k_yy+k_x x)}~,
\]
Where $\psi_k(\xi,\tau)$ satisfies 
\[
\left[
\frac{1}{\xi}\frac{\partial}{\partial\xi} \xi \frac{\partial}{\partial\xi}
-\frac{1}{\xi^2}\frac{\partial^2}{\partial\tau^2}+k^2
\right] \psi_k(\xi,\tau)=0
\]
with $k^2\equiv k_y^2 +k_x^2 +m^2$.
A typical normal mode has the form
\begin{eqnarray}
\psi &=&J_{-i\omega}(k\xi)e^{i\omega\tau} e^{i(k_yy+k_x x)}\\
     &=&\frac{1}{2} \left[ H_{-i\omega}^{(1)}(k\xi)+H_{-i\omega}^{(2)}(k\xi)
                     \right] e^{i\omega\tau} e^{i(k_yy+k_x x)}
\label{eq:typical normal mode}
\end{eqnarray}
Measuring its field consists of sampling it at the events (time $\xi$
and location $\tau$) controlled and calibrated by a set of identically
constructed clocks. If these clocks are inertially expanding clocks as
in Figure~\ref{fig:array of expanding clocks}, then the sampling
events are given by Eqs.(\ref{eq:time of sampling
events})-(\ref{eq:location of sampling events}), and the sampled field
values are
\begin{eqnarray*}
\psi &=&J_{-i\omega}(kb e^{N\Delta\tau})
e^{i\omega M\Delta\tau}e^{i(k_yy+k_x x)}\\
&\propto& (kb)^{-i\omega}e^{-i\omega(N\Delta\tau)}
e^{i\omega M\Delta\tau}e^{i(k_yy+k_x x)}\textrm{ as }N\to -\infty
\end{eqnarray*}
If the samples are are close enough (i.e. $\Delta\tau$ small enough),
then, using Shannon's sampling theorem, one reconstructs the field
from the sampled values of its field.

Note that even though this clock-controlled sampling measurement
reconstructs the the field uniquely in the distant past ($N\to
-\infty$) of $F$, it is clear that this is not the case in the distant
future ($N\to \infty$).  Regardless how small one makes the separation
between the sampling events in the asymptotic past, in the asymptotic
future the inertially expanding clocks tick at such a slow rate (compared
to any atomic clock) that there is no possibility of reconstructing
the field from the sampling measurements.  Indeed, in the distant
future ($\xi=be^{N\Delta\tau}\to\infty$), the field,
Eq.(\ref{eq:typical normal mode}),
\begin{eqnarray*}
\psi\approx\frac{1}{2}\sqrt{\frac{2}{\pi k\xi}}
&~&\!\!\!\!\!\!\left[ e^{-\pi\omega/2}e^{i(k\xi-\pi/4)}+e^{\pi\omega/2}e^{-i(k\xi-\pi/4)}
\right]
\\
&~&\times ~e^{i\omega \tau}e^{i(k_yy+k_x x)}
\end{eqnarray*}
oscillates at a steady rate as a function of (atomic=proper)
$\xi$. But the sampling events, as one can see readily from
Figure~\ref{fig:array of expanding clocks}, are so sparsely spaced as
$N\to\infty$ that there is more than one oscillation between
them. Consequently, reconstruction becomes non-unique and hence out of
the question. In particular, sampling measurements controlled by an
inertially expanding clock are incapable of distinguishing normal
modes traveling into opposite directions, to say nothing about
identifying their oscillation frequencies in the distant future of
$F$.

Would an atomic clock do better? The answer is yes. But only for
sampling measurements made in the distant future
($\xi=be^{N\Delta\tau}\to\infty$). For the distant past
($N\to\infty,~\xi=be^{N\Delta\tau}\to 0$) atomic clocks are just as useless
as inertially expanding clocks are for the distant future: the clocks
simply do not sample the field fast enough to identify its boost
oscillation frequency.

Thus neither atomic clocks nor inertially expanding clocks can give
measurements which identify the nature of the field in both the
asymptotic past and the asymptotic future of $F$. One can measure the
field in one or the other but not both.

A claim that in boost-invariant sector $F$ a pure positive boost
frequency ($\omega$) mode evolves into a superposition of positive and
negative inertial frequency ($\pm k$) modes is wrong. This is because
it makes the tacit assumption that one change inertially expanding to
static atomic clocks in midstream.  Making such a change would go
counter to the central principle of measurement (Section
\ref{sec-Principles of Measurement}): ``once a standard has been
chosen it becomes immutable for all subsequent
measurements''. Violating it would make a standard into a
non-standard.

But a standard is precisely what is needed, otherwise there would be
no way of assigning a frequency and a direction of propagation to
normal modes, the key ingredients to mode amplification and hence to
particle creation as formulated in quantum field theory. Put
differently, an assertion that a mode having a positive frequency evolve
mathematically into a mixture of positive and negative frequency modes
must be accompanied by a specification of a (system of commensurable)
standard clock(s).

It is evident that in sector $F$ no such standard exists.
Consequently, one is not entitled to claim that mathematical analysis
of free fields in that sector predicts the creation of particles.

\section{ACKNOWLEDGEMENTS}
The author appreciates useful conversations about laser physics with
Mark Walker and Linn Van Woerkom.
\bibliography{rindlerspacetime}
\end{document}